\documentclass[a4paper,fleqn,usenatbib]{mnras}

\usepackage{newtxtext,newtxmath}

\usepackage[T1]{fontenc}
\usepackage{ae,aecompl}

\usepackage{graphicx}	
\usepackage{subfigure}
\usepackage{mathtools}
\usepackage{amsmath}	
\usepackage{color}
\usepackage{systeme}
\usepackage{ulem}
\usepackage{booktabs}

\DeclareMathOperator{\sech}{sech}

\definecolor{purple}{rgb}{0.5, 0.0, 0.5}

\renewcommand{\vec}[1]{\mathbf{#1}}

\title[...]{On the Impact of the Numerical Method on Magnetic Reconnection and Particle Acceleration - I. The MHD case.}

\author[E. Puzzoni et al.]{
E. Puzzoni$^{1}$\thanks{E-mail: eleonora.puzzoni@edu.unito.it}, A. Mignone$^{1}$ and G. Bodo$^{2}$
\\
$^{1}$Physics Department, Turin University, Via Pietro Giuria 1, 10125 Torino, Italy\\
$^{2}$INAF – Osservatorio Astrofisico di Torino, Strada Osservatorio 20, I-10025 Pino Torinese, Italy
}

\date{Accepted XXX. Received YYY; in original form ZZZ}

\pubyear{2015}

\begin{document}
\label{firstpage}
\pagerange{\pageref{firstpage}--\pageref{lastpage}}
\maketitle

\begin{abstract}

\noindent We present 2D MHD numerical simulations of tearing-unstable current sheets coupled to a population of non-thermal test-particles, in order to address the problem of numerical convergence with respect to grid resolution, numerical method and physical resistivity.  
Numerical simulations are performed with the PLUTO code for astrophysical fluid dynamics through different combinations of Riemann solvers, reconstruction methods, grid resolutions at various Lundquist numbers.
The constrained transport method is employed to control the divergence-free condition of magnetic field.
Our results indicate that the reconnection rate of the background tearing-unstable plasma converges only for finite values of the Lundquist number and for sufficiently large grid resolutions.
In general, it is found that (for a $2^{\rm nd}$-order scheme) the minimum threshold for numerical convergence during the linear phases requires the number of computational zones covering the initial current sheet width to scale roughly as $\sim \sqrt{\bar{S}}$, where $\bar{S}$ is the Lundquist number defined on the current sheet width.
On the other hand, the process of particle acceleration is found to be nearly independent of the underlying numerical details inasmuch as the system becomes tearing-unstable and enters in its nonlinear stages.
In the limit of large $\bar{S}$, the ensuing power-law index quickly converge to $p\approx 1.7$, consistently with the fast reconnection regime.
\vspace{4mm}

\noindent {\bf Key words:} magnetic reconnection - acceleration of particles - (magnetohydrodynamics) MHD - instabilities - plasmas - methods: numerical
\vspace{4mm}
\end{abstract}

\section{Introduction}
\label{sec:intro}
%
%
The study of the dynamics of  high energy plasmas is of utmost relevance for the interpretation of the phenomenology of high energy astrophysical sources capable of releasing powerful electromagnetic radiation from the radio to the optical, X-ray, and $\gamma$-ray wavebands such as  as blazar jets, \citep[see][]{Bottcher2007, Giannios2013}, gamma-ray bursts \citep[GRBs; see, e.g.,][]{Giannios2008, McKinney2012,Beniamini2014,Beniamini2017}, pulsar wind nebulae \citep[PWNe; see, e.g.,][]{Bucciantini2011,Cerutti2014, Kargaltsev2015,Olmi2016}, and supernova remnants \citep[SNRs; see, e.g.,][]{Amato2009,Morlino2013,Caprioli2014}, among others. 

In these astrophysical environments the electromagnetic field is a key component because, on the one hand it drives, accelerates and partially collimates relativistic outflows from astrophysical black holes, neutron stars and their accretion disks while, on the other hand, magnetic reconnection and dissipation are thought to be responsible for bright thermal and non-thermal emission from these flows \citep{Komissarov2007}. 
During the magnetic reconnection process, fields of opposite polarities rapidly annihilate and the magnetic energy is converted into kinetic and thermal energy of the plasma becoming, at the same time, available in creating a non-thermal population of accelerated particles.
For this mechanism to operate, the plasma cannot be ideal and the Alfv{\'e}n theorem does not hold: field lines should be able to change their topology. 

Albeit astrophysical plasmas are essentially ideal, the flow evolution may lead to the formation of localized regions of large gradients and electric currents, where resistivity cannot be any longer neglected since its role becomes essential in the energy and momentum balance \citep{Mignone2019}.  
The formation of strong and localized current sheets favours reconnection events during which the magnetic field topology may break becoming favourable to resistive instabilities such as the tearing one.
The Sweet-Parker model describes magnetic reconnection although it predicts reconnection rates that are several orders of magnitude slower than the observed ones, especially when the Lundquist number $S = L v_A/\eta$ is large (here $L$ is the characteristic length of the system, $v_A$ is the Alfv\'en velocity and $\eta$ is the resistivity). 
It is in fact a well-known fact that, for astrophysical or laboratory highly-conducting plasmas, $S \gg 1$ (e.g., $S= 10^{12}$ in the solar corona) and classical steady-state models fail to predict the observed bursty phenomena such as solar flares or tokamak disruptions, which occur instead on a non-negligible fraction (say not less than $\sim 10 \%$) of the ideal (Alfv\'enic) timescales \citep[see, e.g,][]{Marcowith2020}. 
A different picture emerges, however, when magnetic reconnection occurs as a time-dependent and unstable process triggered by the tearing mode instability.
In this case, an initially neutral layer tends to fragment into a number of X-points (where the magnetic field has null point) and O-points (or plasmoids) corresponding to regions of high current density. 
Following \cite{Loureiro2016}, it is now generally accepted that the tearing instability radically changes MHD reconnection, that becomes intrinsically time-dependent, bursty and fast compared to the Sweet-Parker steady-state model.
The precise criteria for the onset of the fast reconnection regime, in which the tearing instability occurs basically on the Alfv{\'e}nic timescales, has been numerically investigated in various regimes by \cite{Landi2015} (MHD), \cite{DelZanna2016} (relativistic MHD), and \cite{Papini2019} (Hall-MHD).
We also note that, in addition to the tearing instability, other MHD instabilities triggering turbulence may be equally or more effective in inducing fast magnetic reconnection \citep[as originally claimed by][]{Lazarian1999}, such as the current driven kink instability \citep[CDKI; see, e.g.,][]{Singh2016,Striani2016,Tchekhovskoy2016,Kadowaki2021} or the Kelvin–Helmholtz \citep[KH; see, e.g.,][]{Kowal2020}.

In this respect, the last decade has provided a wealth of investigation, mostly through particle in cell (PIC) numerical simulations, indicating relativistic magnetic reconnection as a very promising candidate in the process of particle acceleration. 
Particle energization via magnetic reconnection can either occur by a direct acceleration in electric fields in the current sheet, or by the anti-reconnection electric field due to the merging of plasmoids \citep[see, e.g.,][]{Oka2010, Sironi2014, Nalewajko2015}, or because of Fermi $1^{\rm st}$-order acceleration \citep[first proposed and quantified by][]{deGouveia2005} in the plasma converging towards the reconnection zone or if particles are trapped in a contracting magnetic islands \citep[see, e.g.,][]{Guo2014, Guo2015, Drake2010, Petropoulou2018, Hakobyan2020}. \cite{Sironi2014}, and likewise  \cite{Petropoulou2018}, found that the accelerated particles populate a power-law distribution with a spectral slope $p = -d \log N / d \log \gamma \sim 2$ for magnetizations $\sigma=10$ in a pair plasma. 

Such studies have been widened to the fluid regime using MHD in conjunction with a test-particle approach \citep[see, e.g.,][]{Liu2009, Gordovskyy2010a, Gordovskyy2010b, Kowal2011, deGouveia2015, Ripperda2017a, Ripperda2017b, Ripperda2019}, finding results similar to those obtained with PIC simulations in terms of particle acceleration.
This has been suggested, e.g., by \cite{Kowal2011}, who have found that in 2D MHD models during an island contraction, in this case due to the merger with other islands, a particle trapped in it can accelerate and increase its energy exponentially in a non-relativistic scenario, which is similar to what \cite{Drake2010} have found with a PIC approach. 
This bulk of evidence suggests that magnetic reconnection can be more efficient and universal when compared to other mechanisms, such as varying magnetic fields in compact sources, stochastic second-order Fermi process in turbulent interstellar and intracluster media, and the first order Fermi process behind shocks \citep{Kowal2011}.  

The fluid plus test-particles approach offers the advantage of being applicable to larger scales when compared to the restrictions imposed by typical PIC simulations, such as resolve the electron skin depth which, in most cases, is several orders of magnitude smaller than the overall size of the astrophysical system, and therefore it would be too expensive in terms of computational cost \citep[see, e.g.,][]{Bai2015, Mignone2018}.
Test-particles acceleration in MHD is usually studied by using frozen snapshots \citep[see, e.g.,][]{Liu2009, Gordovskyy2010a, Kowal2011, deGouveia2015, Ripperda2017a} in which the fluid provides a background, static configuration on top which particles are allowed to evolve.

With the exception of very few works \citep[see, e.g.,][]{Kowal2009, Santos2010}, however, the vast majority of these studies largely overlooks the impact of the numerical method on the simulation results, it scarcely addresses the problem of convergence with respect to grid resolution and it often neglects the effect of a physical resistivity on the evolution of the instabilities.

Classical and relativistic MHD numerical simulations have, in fact, shown that when the Lundquist number $S$ is greater than a certain threshold value, plasma instabilities in the current sheet trigger a fast reconnection regime with time-scales comparable to the observed ones. 
However, the grid resolution must be sufficiently large to ensure that the dissipation scale is regulated by physical resistivity and not by numerical diffusion. 
This requirement together with the low-resistivity (or high S, typically $S > 10^3$) typical of astrophysical plasma can indeed demand very fine mesh spacing in proximity of the current sheet.
This determines the numerical convergence of the simulation and eventually regulates the correct reconnection rate, once proper numerical resolution is achieved. 

Based on these considerations, in this work we intend to address a number of unresolved issues by assessing the impact of numerical method, grid resolution and physical resistivity on the magnetic reconnection process as well as their implications in the particle acceleration mechanisms.
We will start by considering a test-particle approach using 2D non-relativistic MHD numerical computations carried out with different Riemann solvers, spatial reconstruction both in the ideal and resistive MHD regimes at different values of the Lundquist number.
The relativistic extension will be considered in a companion paper.
In addition and contrary to most previous investigations, test-particles and MHD fluid will be evolved simultaneously with the advantage of studying the acceleration mechanism in response to the dynamical evolution of the system, as done by \cite{Gordovskyy2010b}, \cite{Kowal2012} and \cite{Ripperda2017b, Ripperda2019}.  

The paper is organized as follows. 
The classical MHD equations describing the evolution of the fluid and the particle equations of motion are discussed in Section \ref{sec:model}, along with the numerical setup. 
The results obtained for the fluid case only, and which therefore include the comparison between the spatial reconstructions and the Riemann solvers, are shown in Section \ref{sec:results_fluid}.
Instead, the results obtained for the particles, including the impact on particle acceleration properties of the grid resolution and resistivity, are shown in Section \ref{sec:results_part}. 
A summary is given in Section \ref{sec:summary}.

\section{Relevant Equations and Model Setup}
\label{sec:model}
%
%

\subsection{The Resistive MHD Equations}
%

Magnetohydrodynamics (MHD) describes an electrically conducting single fluid, assuming low frequency and large scales.
MHD provides indeed the basic description of a plasma at the macroscopic level by neglecting kinetic effects and electron physics, approximations commonly used for applications to laboratory, space and astrophysical plasmas. 
The resistive non-relativistic MHD equations, which include the continuity, momentum, induction and energy conservation laws are, respectively,

\begin{align}
  \frac{\partial \rho}{\partial t} & + \nabla\cdot(\rho\textbf{v}_g) = 0 \,,
  \label{eq:continuity} \\ \noalign{\smallskip}
  \frac{\partial \textbf{m}}{\partial t} &+ \nabla\cdot
  \left[\textbf{m} \textbf{v}_g - \textbf{B}\textbf{B} + \mathrm{I}\left(p + \frac{\textbf{B}^2}{2}\right)\right] =
   0 \,,
  \label{eq:momentum} \\ \noalign{\smallskip}
  \frac{\partial \textbf{B}}{\partial t} &+ \nabla \times (c\textbf{E}) = 0  \,,
  \label{eq:induction} \\ \noalign{\smallskip}
  \frac{\partial E_t}{\partial t} & + \nabla\cdot \left[\left(\frac{\rho \textbf{v}^2_g}{2} + \rho e + p \right)\textbf{v}_g + c\textbf{E} \times \textbf{B}\right] = 0   \,,
  \label{eq:energy}
\end{align}
where $\rho$ is the mass density, $\textbf{m} = \rho \textbf{v}_g$ is the momentum density, $\textbf{v}_g$ is the gas velocity, $p$ is the gas (thermal) pressure, $\textbf{B}$ is the magnetic field (a term $\sqrt{4 \pi}$ is included in the definition of $B$) and $E_t$ is the total energy density:
\begin{equation}
  E_t = \rho e + \frac{\textbf{m}^2}{2 \rho} + \frac{\textbf{B}^2}{2},
\end{equation}
where, for an ideal gas, we have $\rho e = p/(\Gamma-1)$.
The relative importance of thermal (gas) and magnetic pressures is quantified by the $\beta$ parameter, defined as
\begin{equation}
  \beta = \frac{p}{B^2/ 2}.
\end{equation} 
If $\beta \ll 1$ the magnetic field dominates the dynamics while, in the opposite limit, gas motion drags the field lines so that the magnetic field behaves essentially in a passive way.

The magnetic field evolution is governed by Faraday's law (Eq. \ref{eq:induction}) where $\textbf{E}$ is the electric field defined by
\begin{equation}\label{eq:electricfield}
  c \textbf{E} = - \textbf{v}_g \times \textbf{B} + \frac{\eta}{c} \textbf{J},
\end{equation}
where, the first term in Eq. (\ref{eq:electricfield}) is the convective term while the second one corresponds to the resistive electric field (i.e $\eta$ denotes the scalar resistivity) with the current density defined as 
\begin{equation}
  \textbf{J} = c\nabla \times \textbf{B}.
  \label{eq:currentdensity}
\end{equation}
The presence of the resistive term in Eq. (\ref{eq:electricfield}) is of crucial importance in triggering tearing-driven magnetic reconnection as well as in the process of particle acceleration \citep{Li2017}. 
While in ideal MHD no electric field is present in the fluid rest frame, a resistive plasma is still capable of accelerating particles at stagnation points, provided a large current is formed.
In reconnecting current sheet this condition is manifestly evident at X-points where the condition $|\vec{E}| > |\vec{B}|$ can easily occur. 

Eq. (\ref{eq:electricfield}) introduces two time scales, namely, the convective one ($\tau_c = L/v_g$), and the diffusive one ($\tau_d = L^2/\eta$), where $L$ represents the typical system length scale. 
The relative importance of the two terms can be quantified by the magnetic Reynolds number, defined by the ratio of these, $R_m = \tau_d / \tau_c = v_g L/\eta$ or, more conveniently, by the Lundquist number
\begin{equation}\label{eq:S}
  S = \frac{v_A L}{\eta}
\end{equation}
where $v_A$ denotes the Alfv\'en speed
\begin{equation}
  v_A = \frac{B}{\sqrt{\rho}}, 
\end{equation}
While the magnetic field topology is generally preserved in the ideal MHD regime ($S\to\infty$) owing to the frozen-in condition, this is not the case for finite values of $S$ and diffusion of magnetic field is possible. 
Indeed, even if astrophysical plasmas are typically highly conductive (e.g. $S\gg 1$), fast magnetic reconnection \citep{Giannios2013, DelZanna2016} can become an effective process in localized dissipation regions featuring very thin current sheets. 
This process is ultimately driven by the tearing instability and is responsible for their fragmentation into a large number of plasmoids. 

Finally, Faraday's law is complemented with the solenoidal condition of magnetic field, 
\begin{equation} \label{eq:solenoidal}
  \nabla \cdot \textbf{B} = 0,
\end{equation}
so that, if true initially, it must be preserved during the subsequent time evolution. 

\subsection{Particle Equations of Motion}
%

Particles are defined in terms of their spatial coordinates $\textbf{x}_p$ and velocity $\textbf{v}_p$, which are governed by the equation of motion
\begin{equation}
\label{eq:particles}
\begin{dcases} 
\frac{d \textbf{x}_p}{d t} = \textbf{v}_p \\ 
\frac{d (\gamma \textbf{v})_p}{d t} = \left(\frac{e}{mc}\right)_p (c \textbf{E} + \textbf{v}_p \times \textbf{B})
\end{dcases}
\end{equation}
where $\gamma= 1/\sqrt{1 - \textbf{v}^2_p/\mathbb{C}^2}$ is the Lorentz factor while $(e/mc)_p$ is the particle charge to mass ratio. The suffix $p$ will be used to label a single particle. 
In the MHD equations the actual speed of light does not explicitly appears, therefore the artificial value $\mathbb{C}$ is used. 
In this paper we have set $\mathbb{C} = 10^4 v_A$ (where $v_A$ is the Alfv{\'e}n speed) since, for  consistency  reasons, it must be much larger than any characteristic signal velocity. 
The electric and magnetic fields $\textbf{E}$ and $\textbf{B}$ are computed from the magnetized fluid and are properly interpolated at the particle position, following the approach of \cite{Mignone2018}.

We solve the particle equations of motion using the Boris integrator, which is time reversible and features good conservation properties for long time simulation. 
For computational efficiency, the particle mass is taken to be equal to the mass of the particles composing the fluid so that, when written in code units, the charge to mass ratio in Eq. (\ref{eq:particles}) becomes unity.
In other words, our results are equally applicable to protons embedded in protons+electron thermal fluid or to a electron-positron pair plasma.

As discussed in \cite{Mignone2018}, the particle time step can be constrained either by (the inverse of) their gyration frequency or by the maximum number of computational zones that can be crossed during a single time step.
In our simulations, both are found to be smaller than the fluid time step by a factor of $\sim 5$ at nominal resolutions.
While gyration dominates the time step restriction at low grid resolutions, the opposite situation is found as the mesh becomes finer, since the most energetic particles can cross increasingly more cells in a single time step.

\subsection{Initial and boundary conditions}
\label{sec:numerical_setup}
%

Our initial configuration considers a 2D rectangular domain of size $L \times L/2$ where $L=2\mathbb{C}/\omega_p = 2\times10^4 \ v_A/\omega_p$ and $\omega_p$ represents the plasma frequency, which naturally arises when Eq. (\ref{eq:particles}) is scaled-down to the MHD equations.
The equilibrium magnetic field follows a Harris-sheet profile,
\begin{equation}\label{eq:HarrisB}
  B_x(y) = B_0 \tanh \left(\frac{y}{a}\right),
\end{equation}
where $a$ denotes the initial width of the current sheet, which we set to $a = 250 \ v_A/\omega_p$ in all the simulations, while $B_0$ denotes the magnetic field strength, normalized such that our unit velocity is the Alfv\'en speed ($\rho_0 = B_0 = 1$ in code units).

The guide field is not present ($B_z=0$) and an initial equilibrium condition is obtained by counteracting the Lorenz-force term with a thermal pressure gradient,
\begin{equation}
  p(y) = \frac{1}{2} B_0^2 (\beta + 1) - \frac{1}{2}B^2_x(y)\,,
  \label{eq:equilibrium}
\end{equation}
so that the total pressure remains constant through the sheet.
In all simulations we use $\beta = 0.01$.
For convenience, resistivity is prescribed from $\bar{S}$, i.e., Lundquist number corresponding to the current sheet width,
\begin{equation}\label{eq:Sbar}
  \bar{S} = \frac{aS}{L} = \frac{v_A a}{\eta}
\end{equation}
where $v_A$ is set to one.
Boundary conditions in the $x$-direction are periodic, while in the $y$-direction are reflective.

The system is perturbed by introducing a fixed number of small-amplitudes modes with different wavenumbers $k$.
This is best achieved by redefining the vector potential as $A_z(x,y) = A_{0}(y) + \delta A_z(x,y)$ where $A_0(y) = a B_0 \log(\cosh(y/a))$ corresponds to the equilibrium field (Eq. \ref{eq:HarrisB}) while
\begin{equation}
  \delta A_z(x,y) = \frac{\epsilon B_0}{N_m}
               \sum_{m=0}^{N_m} \frac{1}{k}\sin(kx + \phi_m)                                \sech\left(\frac{y}{a}\right)
\end{equation}
is the perturbed term, $N_m$ is the number of modes ($20$ in our case), $\epsilon = 10^{-3}$ is the perturbation amplitude, $\phi_m$ are random phases and $k=(m+1)k_0 = 2\pi (m+1)/ L $. 

Unless otherwise stated, we choose our final integration time as $\omega_p t_f = 6 \cdot 10^5$, enough for all models to capture both the linear stages as well as the nonlinear evolution. 
Notice that, for convenience, time can also be measured in units of the Alfv{\'e}n time-scale $\bar{\tau}_A=a/v_A=250/\omega_p$.
In practice, the corresponding Alfv{\'e}n time can be obtained from the simulation time (in units of $1/\omega_p$) as $t/\bar{\tau}_A = \omega_p t/250$.

Test-particles are evenly assigned to grid zones (we use 1 particle / cell) and their velocities follow a Maxwellian distribution with standard deviation $\sigma = \sqrt{p_0/2 \rho_0}$, where $p_0=p(\infty)$ in Eq. (\ref{eq:equilibrium}).

\subsection{Chosen numerical methods}
\label{sec:num_methods}
%

Eqns (\ref{eq:continuity})-(\ref{eq:energy}) are solved numerically using the PLUTO code for computational plasma physics, see \cite{Mignone2007} and \cite{Mignone2012}.
PLUTO employs a finite volume formulation whereby the integral form of the equations is discretized and conserved variables (density, momentum and energy) are evolved in terms of their volume averages (rather than point values) inside the computational zone.
The magnetic field divergence-free condition is satisfied to machine precision using the constrained transport (CT) method, originally introduced by \cite{Evans1988} and later extended to Godunov type schemes by \cite{Balsara1999},  \cite{Londrillo_DelZanna2004}, \cite{Gardiner2005} to name just a few.
CT methods entail to a staggered discretization of magnetic field components so that a discrete version of Stoke's law can be applied when solving the induction equation.
In PLUTO, the line-average electric field (the electromotive force or emf) is constructed using the information available from one-dimensional, face centered Riemann solver and we refer the reader to the recent work of \cite{Mignone2021} for a thorough description of different algorithms employed in the present context.

As we are aiming at quantifying the impact of the numerical scheme on simulation results, we compare three different numerical schemes based on different combinations of the base Riemann solver and the EMF averaging/reconstruction scheme, namely, 
\begin{itemize}
    \item[i)] the HLL Riemann solver with the UCT-HLL reconstruction \citep{Londrillo_DelZanna2004, delZanna_etal2007};
    \item[ii)] the Roe Riemann solver \citep{Cargo1997} with the CT-Contact emf averaging scheme of \cite{Gardiner2005};
    \item[iii)] the HLLD Riemann solver of \cite{MiyKus.2005} with the more recent UCT-HLLD emf averaging scheme;
\end{itemize}
which, for brevity, will be shortened as HLL (i), Roe (ii) and HLLD (iii).
A detailed inter-scheme comparison is presented in \cite{Mignone2021}.
The first two schemes are being ordinarily employed in numerical simulations of MHD flows and differ in the amount of numerical diffusion (the former being more diffusive than the latter).
The third scheme (HLLD) is more recent and presents excellent stability properties and reduced numerical dissipation when applied to time-dependent magnetized current sheets. 

While the code retains a global $2^{\rm nd}$-order accuracy, the amount of numerical dissipation can be further controlled by the spatial reconstruction of fluid variables inside each grid zone.
For this reason we consider both $2^{\rm nd}$-order piecewise linear reconstruction and the $5^{\rm th}$-order WENO-Z algorithm \citep[see, e.g.,][]{Borges2008, Mignone2010}. 
A $3^{\rm rd}$-order Runge Kutta time stepping is used to advance the equations in time.
Test-particles are also evolved together with the fluid using the Boris algorithm.

We conduct numerical simulations using different values of the Lundquist number $\bar{S}$ (Eq. \ref{eq:Sbar}), namely, $\bar{S} = 10^3,\, 10^4,\, 10^5$ and $\bar{S} = \infty$ (ideal case) with varying grid resolutions, starting from $N_x=192$ (which corresponds to $a/\Delta x \sim 2.5$ zones on the initial current sheet width $a$) up to $N_x=3072$ ($a/\Delta x \sim 40$).
Correspondingly, the number of particles varies from 18,432 (at the lowest resolution) and reaches 4,718,592 (at the largest one).

\begin{figure*}
  \centering
  \includegraphics[width=0.45\textwidth]{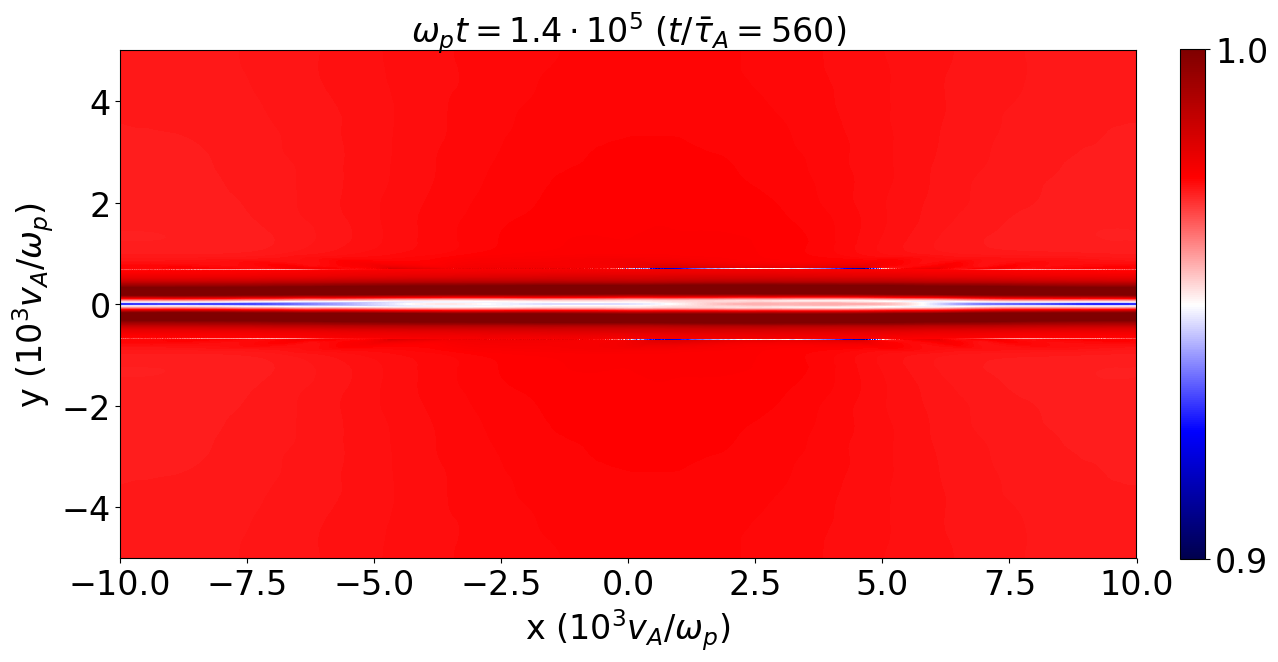} \hspace{0.5pt}
  \includegraphics[width=0.45\textwidth]{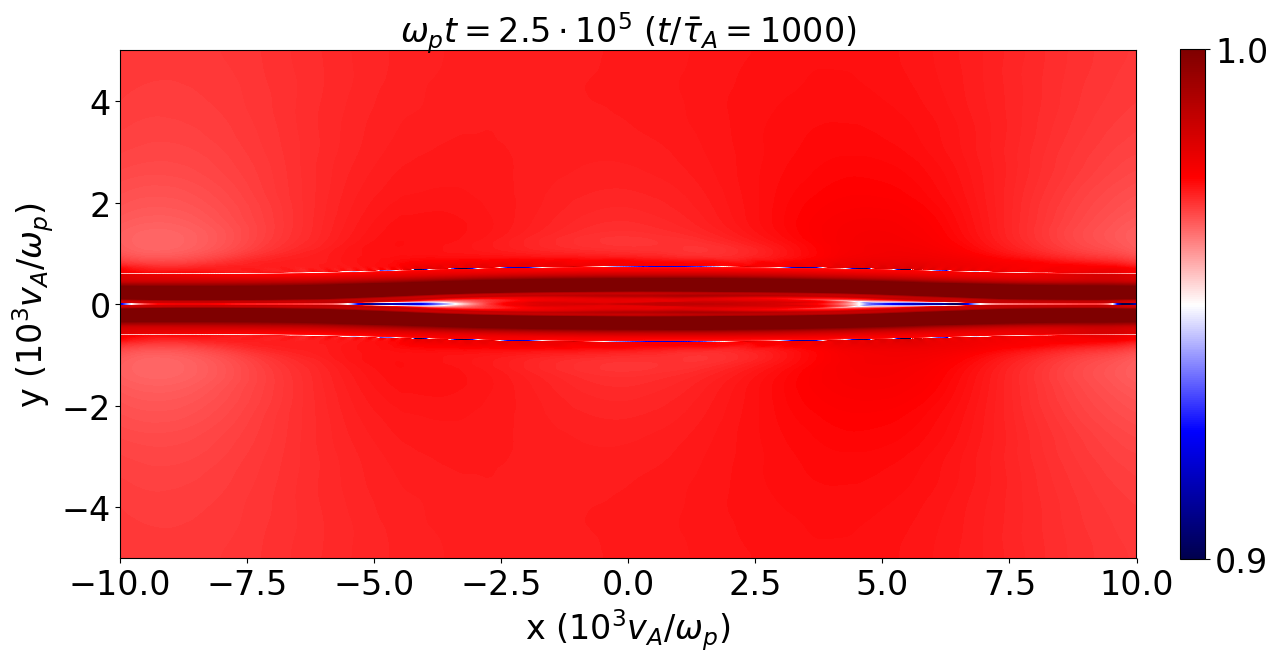} \\
  \includegraphics[width=0.45\textwidth]{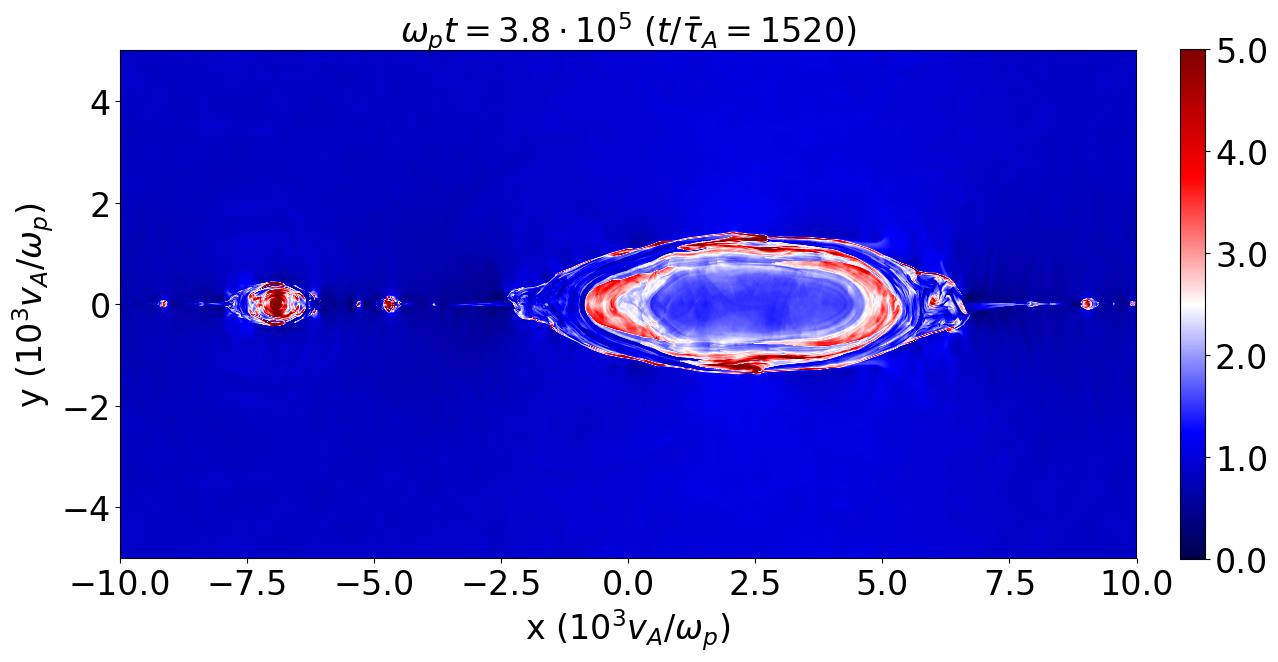} \hspace{0.5pt}
  \includegraphics[width=0.45\textwidth]{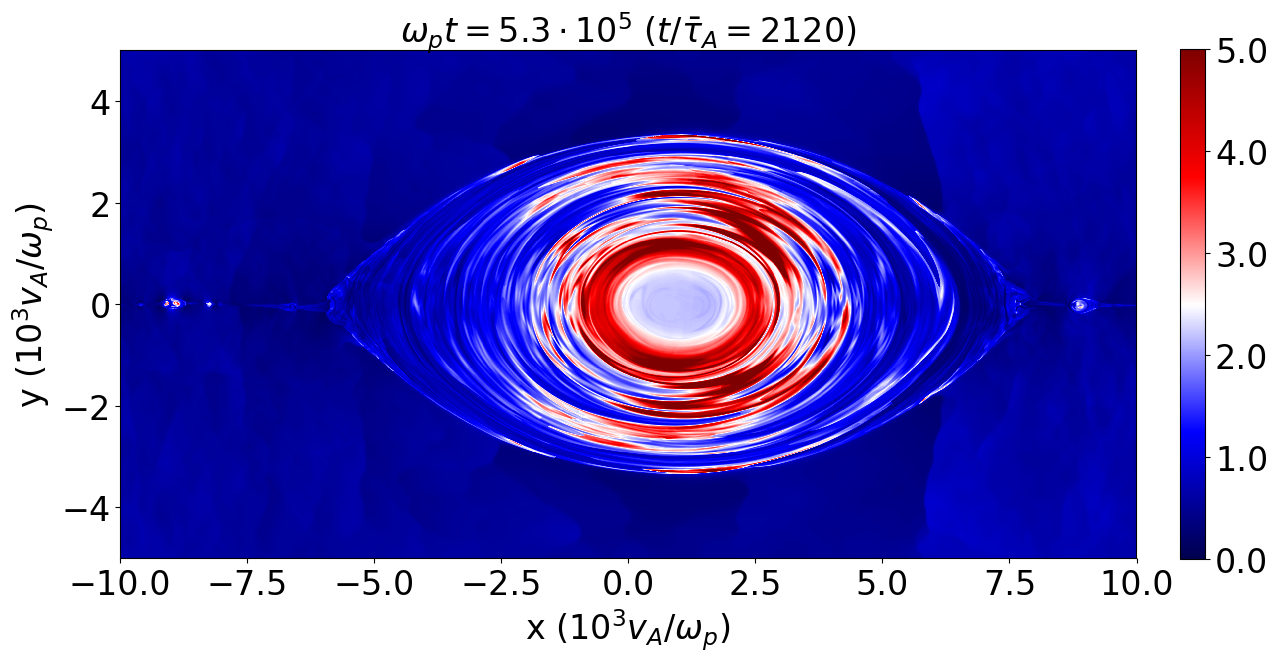}
  \caption{Plasma density (colorbar) snapshots at four simulation snapshots, obtained with a grid resolution $N_x = 3072$, the UCT-HLLD scheme and the HLLD Riemann solver with $\bar{S}=10^4$ and the WENO-Z scheme.
  Time is expressed in units of both the inverse plasma frequency and the Aflv{\'e}nic time scale (in parenthesis).}
  \label{fig:plasmaevolution}
\end{figure*}

\begin{figure*}
  \mbox{\includegraphics[width=0.36\textwidth]{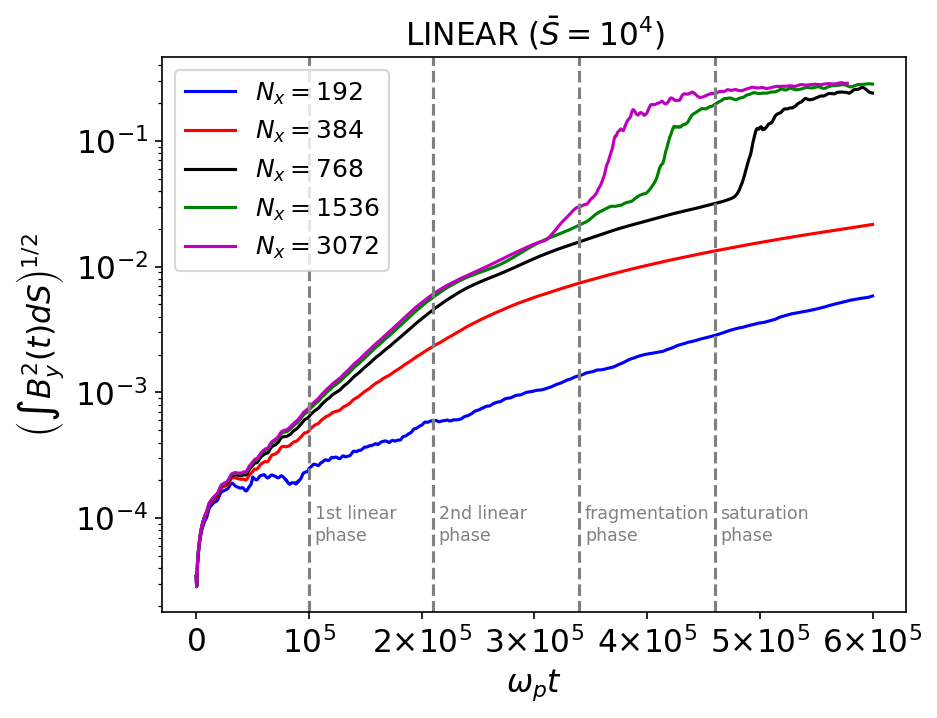}}   
  \mbox{\includegraphics[width=0.31\textwidth]{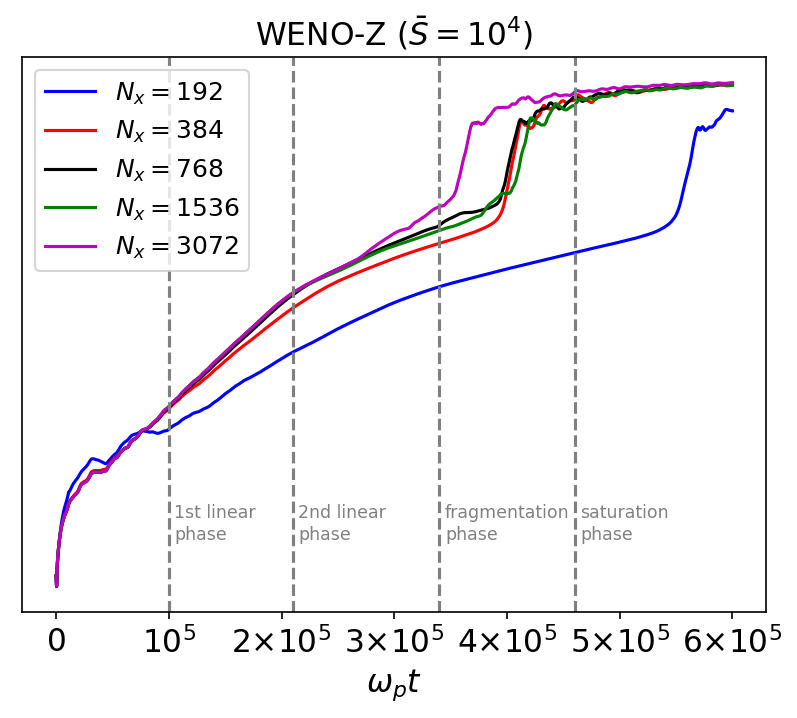}}
  \mbox{\includegraphics[width=0.31\textwidth]{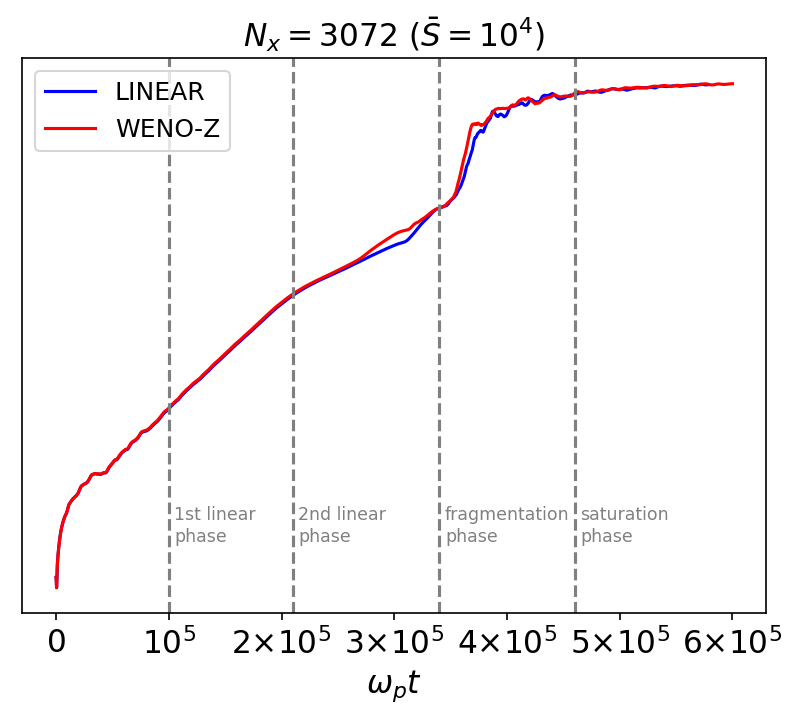}}  
  \mbox{\hspace*{0.07cm} \includegraphics[width=0.35\textwidth]{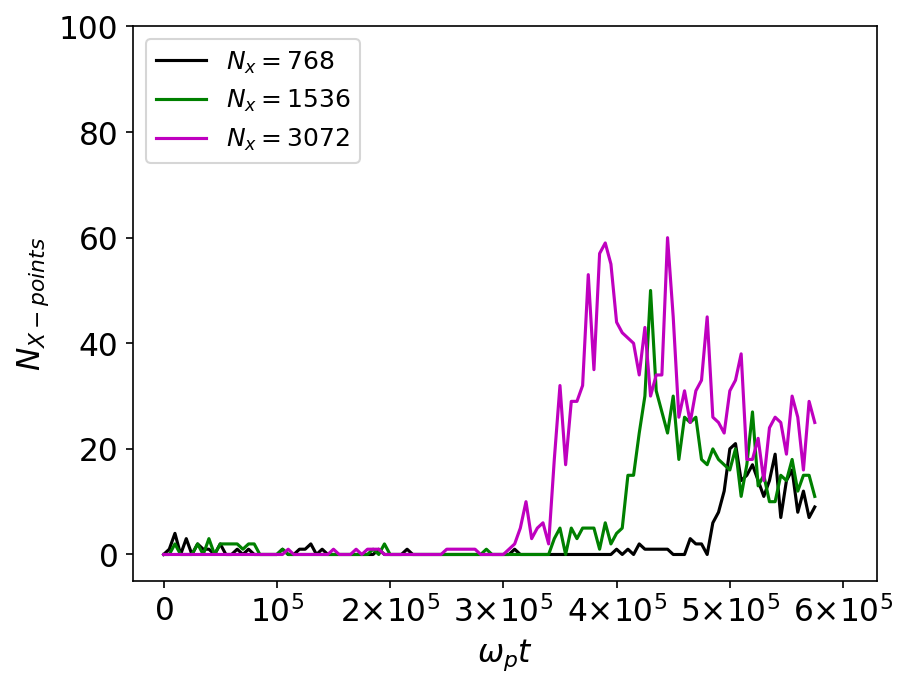}}   
  \mbox{\includegraphics[width=0.31\textwidth]{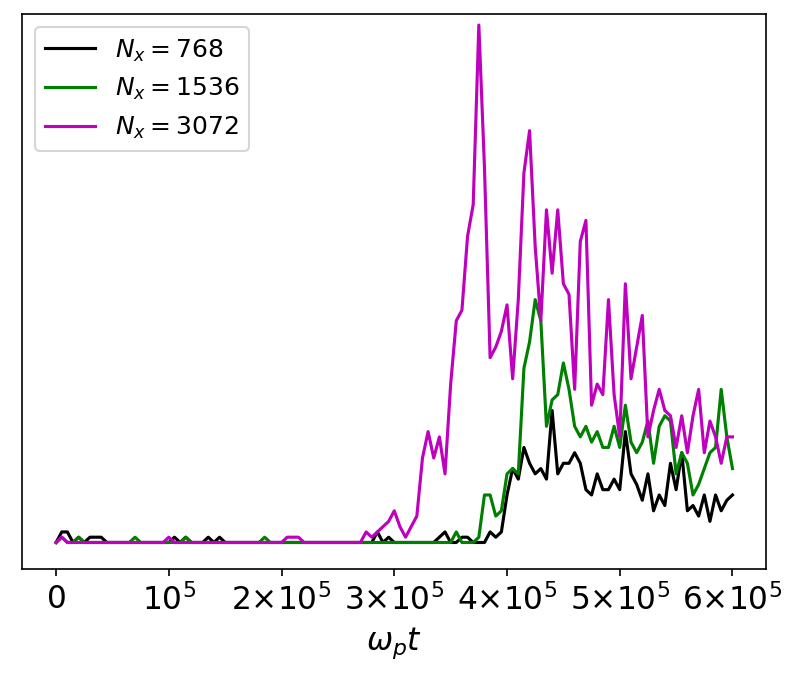}}
  \mbox{\includegraphics[width=0.31\textwidth]{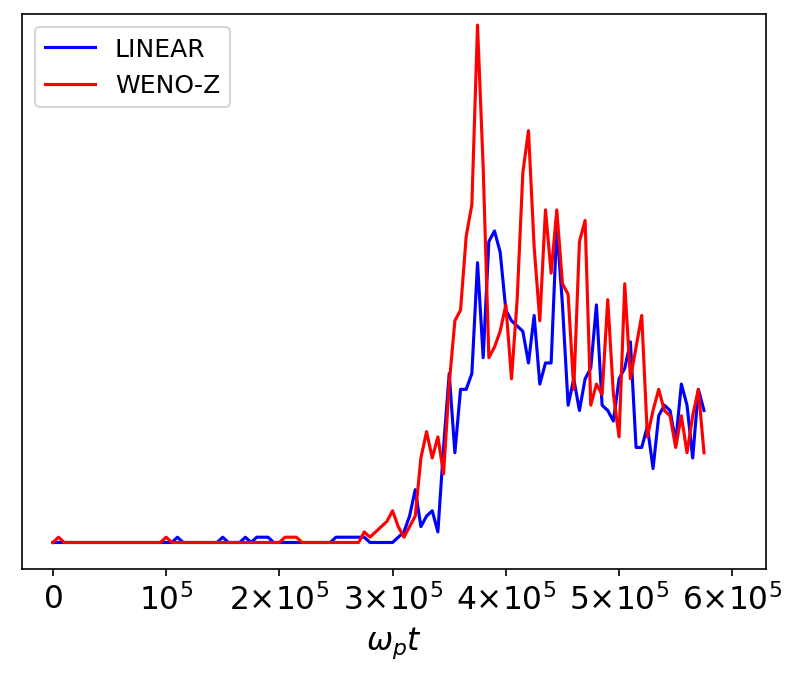}}
  \caption{\textit{Top panels:} spatially-averaged transverse component of magnetic field as a function of time at different grid resolutions, using linear (left panel) and WENO-Z (middle panel) reconstructions. 
  The rightmost panel compares the two reconstruction at the largest resolution $N_x=3072$ (right panel). 
  Note that time is expressed in units of the inverse plasma frequency and it can be converted to Alfv{\'e}nic time units using $t/\bar{\tau}_A = \omega_p t/250$.
  \textit{Bottom panels}: number of X-points formed over time at higher resolutions using linear (left panel) and WENO-Z (middle panel) reconstruction with a comparison of these at $N_x = 3072$ (right panel). We used $\bar{S} = 10^4$ and the HLLD Riemann solver with the UCT-HLLD scheme in all cases.
}
  \label{fig:LINEARvsWENOZ}
\end{figure*}

\section{Convergence Study for the Background Plasma}
\label{sec:results_fluid}
%
%

We now focus on the dynamical evolution of the tearing-unstable current sheet and proceed to assess the impact of the numerical method on the simulation results.

Figure \ref{fig:plasmaevolution} shows the temporal evolution of the plasma density obtained with the HLLD scheme and $\bar{S}=10^4$ at the largest grid resolution ($N_x=3072$).
For $10^5 \lesssim \omega_p t \lesssim 2.1 \cdot 10^5$ (upper left panel) the current sheet  starts to shrink at the edges of the computational domain and the tearing-mode heads its linear phase.
The process continues as the current sheet becomes thinner and thinner for $2.1 \cdot 10^5 \lesssim \omega_p t \lesssim 3.4 \cdot 10^5$ (upper right panel), therefore we call this phase the second linear phase. 
Subsequently, at about $3.4 \cdot 10^5 \lesssim \omega_p t \lesssim 4.6 \cdot 10^5$ the current sheet starts to fragment into plasmoids that begin to merge (lower left panel). 
Eventually, for $\omega_p t \gtrsim 4.6 \cdot 10^5$, the system reaches a saturation phase where plasmoids merged into a single large island with a large filling factor (lower right panel). 

Regardless of the numerical scheme, these four evolutionary stages are observed at all resolutions, albeit the beginning of each phase may occur at a different time. 
In our experience, we have found that a convenient way to label the different evolutionary stages can be quantified by counting the number of X-points formed over time.
The algorithm, based on locating the null points of $|\vec{B}|$, is illustrated in Appendix \ref{app:xpoints} and is compatible with the one developed by \cite{Zhdankin2013}.

As an additional diagnostic tool, we also provide a quantitative measure of the growth rate $\gamma_{\textsc{\tiny{\rm TM}}}$, obtained as
\begin{equation}\label{eq:slope}
  \gamma_{\textsc{\tiny{\rm TM}}} = \frac{f(t_2) - f(t_1)}{t_2 - t_1},
\end{equation}
where $t_1$ and $t_2$ correspond to two simulation snapshots while
\begin{equation}\label{eq:Bt}
  f(t) = \frac{1}{2} \log \left(\frac{1}{L^2}\int B_y^2(t)\,dx\,dy\right).
\end{equation}
Eq. (\ref{eq:slope}) is employed to evaluate the growth rate within the $1^{\rm st}$ and $2^{\rm nd}$ linear phases in what follows.
\subsection{The effect of spatial reconstruction}

We plot, in the top panels of Figure \ref{fig:LINEARvsWENOZ}, the temporal evolution of the spatially-averaged transverse component of magnetic field at different resolutions using linear reconstruction (top left) and WENO-Z $5^{\rm th}$-order reconstruction (top middle).
The bottom panels give the corresponding number of X-points at the largest resolutions ($768 \le N_x \le 3072$).
For the sake of comparison, results with both reconstruction schemes at $N_x=3072$ are superimposed in the rightmost panels.

For this case, we set $\bar{S} = 10^4$ and employ the HLLD Riemann solver with the UCT-HLLD reconstruction scheme for EMF computation at cell edges. 
The different evolutionary stages, described above, have been marked by vertical dashed lines.
During the linear phases, we expect perturbations to grow exponentially at the rate of the fastest growing mode, as predicted by linear theory, see \S\ref{sec:Lundquist}.

\begin{table}
	\centering
	\caption{Average growth rates (for $\bar{S}=10^4$ case) for the tearing mode instability, $\gamma_{\textsc{\tiny{\rm TM}}}$, measured from the simulations at different grid resolutions (left column) using HLLD with linear and WENO-Z reconstructions.
	These are calculated within what we call the first ($10^5 < \omega_p t < 2.1 \cdot 10^5$ or, equivalently, $400 < t/\bar{\tau}_A < 840$) and the second linear phase ($2.1 \cdot 10^5 < \omega_p t < 3.4 \cdot 10^5$ or, equivalently, $840 < t/\bar{\tau}_A < 1360$).}
	\label{tab:slopes}
    \centering
	\begin{tabular}{cccccc} 
                \hline
                & & \multicolumn{2}{c}{$1^{st}$ linear phase} & \multicolumn{2}{c}{$2^{nd}$ linear phase} \\
                & & \multicolumn{2}{c}{$\gamma_{\textsc{\tiny{\rm TM}}} \ (10^{-5} \omega_p)$} & \multicolumn{2}{c}{$\gamma_{\textsc{\tiny{\rm TM}}} \ (10^{-6} \omega_p)$} \\
                \cmidrule{3-4}  \cmidrule{5-6}
                Resolution & $a/\Delta x$ & Linear & WENO-Z & Linear & WENO-Z \\
		\hline
		192 $\times$ 96 & $\sim2.5$ & 0.80 & 1.29 & 6.38 & 9.24 \\
		384 $\times$ 192 & $\sim5$ &  1.40 & 1.68 & 8.93 & 9.14 \\
		768 $\times$ 384 & $\sim10$ &  1.76 & 1.88 & 9.58 & 9.80 \\
        1536 $\times$ 768 & $\sim20$ & 1.88 & 1.89 & 10.2 & 8.85 \\
        3072 $\times$ 1536 & $\sim4$ & 1.89 & 1.90 & 12.3 & 12.2 \\
		\hline
	\end{tabular}
\end{table}

\begin{figure*}
  \centering
  \includegraphics[width=0.365\textwidth]{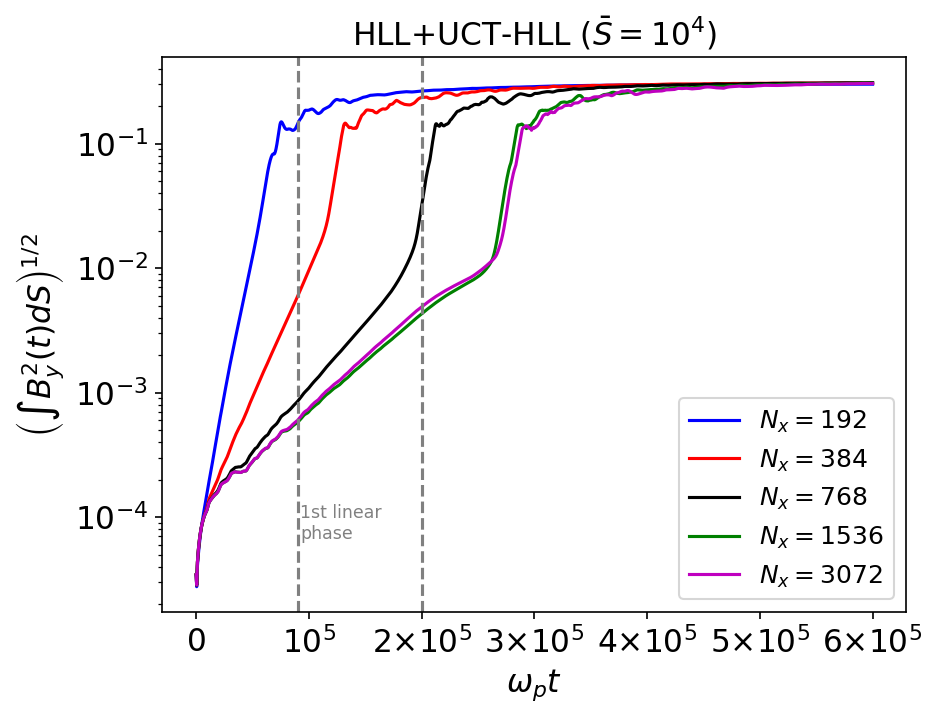} 
  \includegraphics[width=0.312\textwidth]{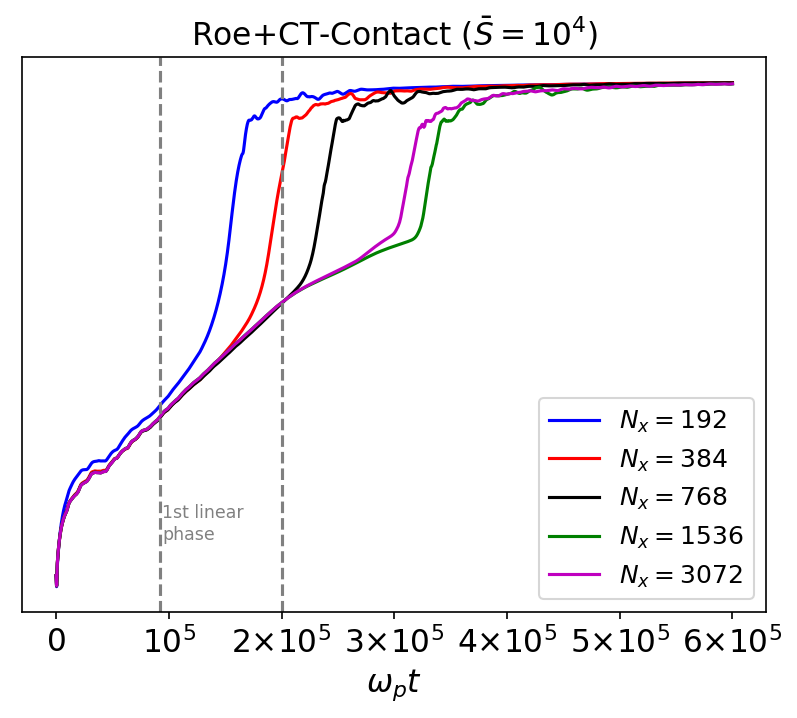}    
  \includegraphics[width=0.313\textwidth]{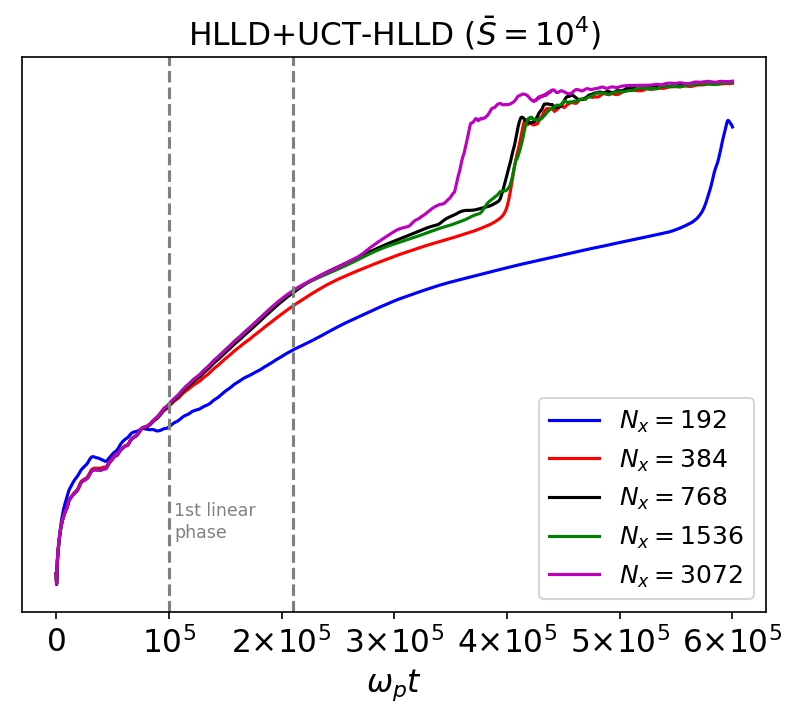}  \\
  \includegraphics[width=0.365\textwidth]{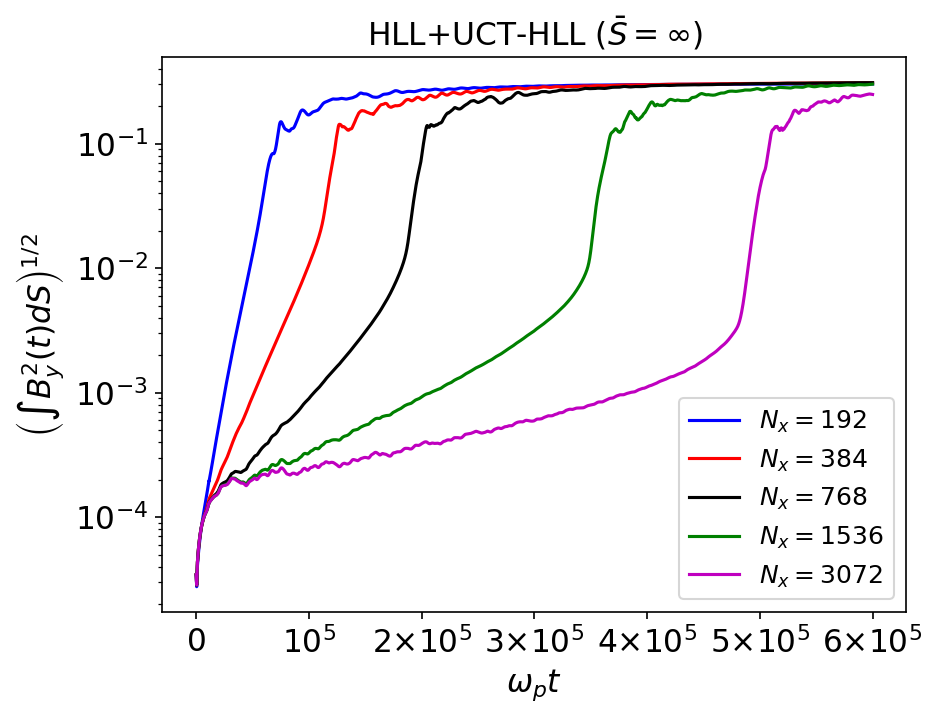} 
  \includegraphics[width=0.312\textwidth]{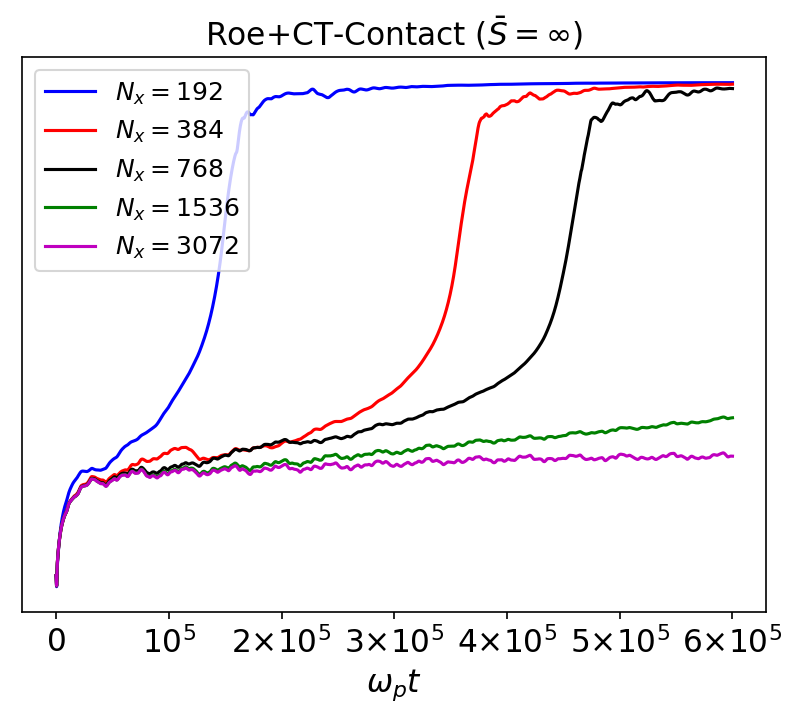}    
  \includegraphics[width=0.313\textwidth]{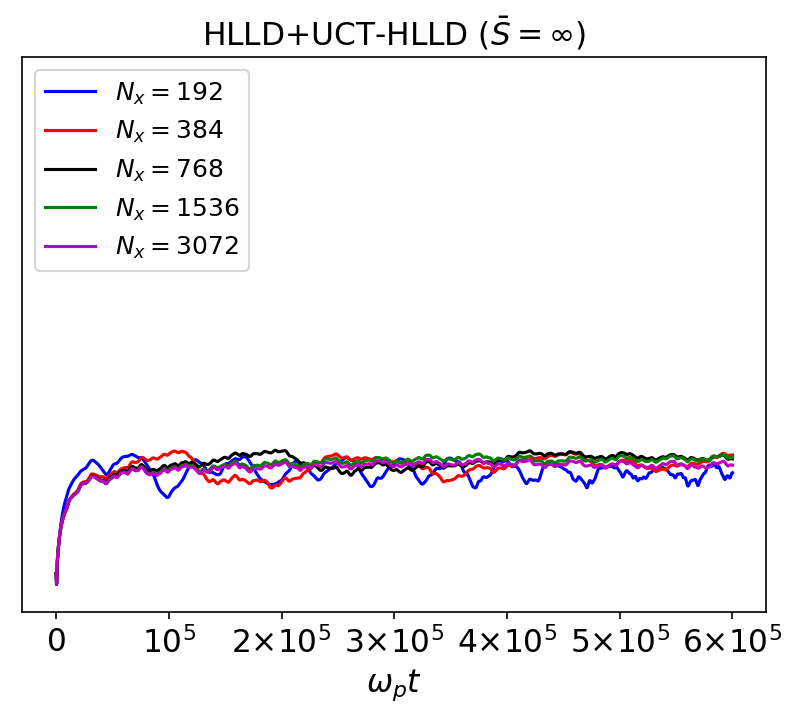}
  \caption{Spatially-averaged transverse component of magnetic field as a function of time for different resolutions and selected numerical schemes, with $\bar{S}=10^4$ (top panels) and $\bar{S}=\infty$ (bottom panels).
  The WENO-Z reconstruction is used for all computations. 
  Convergence refers to the $1^{\rm st}$ linear phase only (larger resolution is needed to resolve smaller current sheets forming after the fragmentation phase).}
  \label{fig:resvsideal}
\end{figure*}

The same does not hold during the subsequent phase where a variation of the magnetic energy is visible depending on the chosen scheme and grid size. In fact, during the second phase the width of the current sheet continues to decrease, eventually
leading to its fragmentation through the formation of X-points. In order to accurately capture this phase, an even larger resolution is needed for resolving the increasingly thinner sheets.

A direct inspection of the top and corresponding bottom panel reveals that the growth of the perturbation raises with the number of newly formed X-points.
Since this process occurs more rapidly as the numerical diffusion is reduced, higher resolution runs with WENO-Z reconstruction evolve towards the growth of saturation earlier. 
On the contrary, at low resolutions, the saturation stage is attained at later times or may not be reached at all by the end of the simulation, specially with linear reconstruction where numerical diffusion is larger. 
Focusing on the linear phases, convergence is eventually reached for $N_x\gtrsim 1536$ ($a/\Delta x \simeq 20$) for the linear reconstruction case and $N_x \gtrsim 768$ ($a/\Delta x \simeq 10$) using WENO-Z. 
At the highest resolution, both linear and WENO-Z schemes show similar growths (see rightmost top panel).

Growth rates for the $1^{\rm st}$ and $2^{\rm nd}$ linear phases, computed using Eq. (\ref{eq:slope}), are reported in Table \ref{tab:slopes}.
During the $1^{\rm st}$ linear stage, both computations eventually converge to the same result, albeit the employment of WENO-Z favours faster convergence (approximately half the resolution is needed), owing the reduced numerical dissipation.
The $2^{\rm nd}$ linear phase takes place more rapidly and starts  earlier as the numerical dissipation is reduced (either with the reconstruction order or with a finer mesh spacing).

\subsection{The impact of the Riemann Solver and emf Averaging}
\label{sec:riemann}

\begin{figure}
  \centering
  \includegraphics[width=0.5\textwidth]{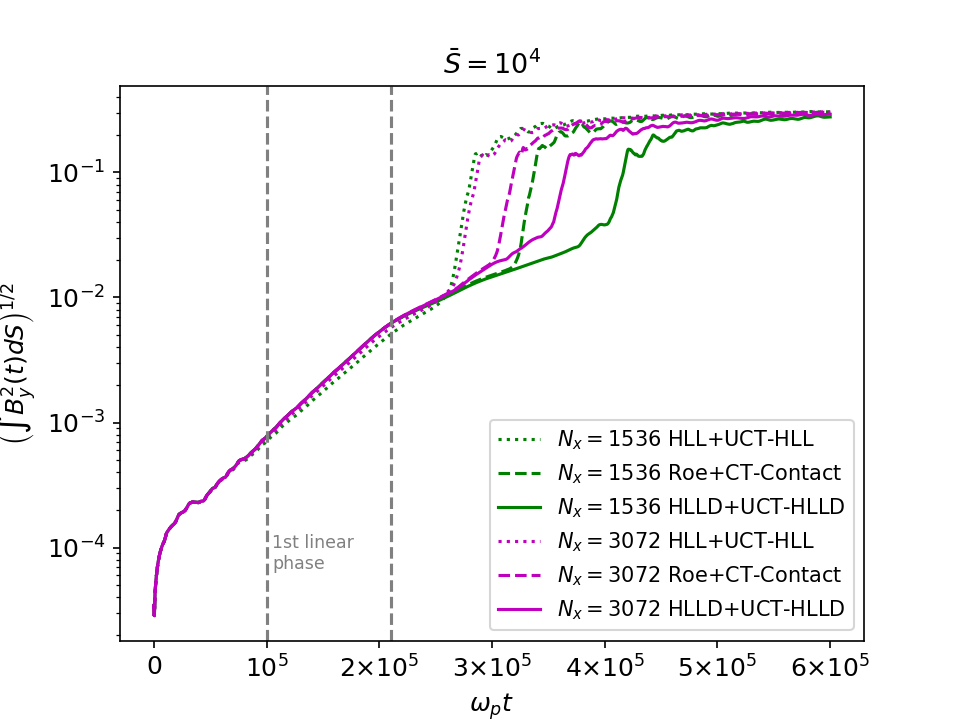}
  \caption{Same as Fig. \ref{fig:resvsideal} but with 1D plots from different numerical methods overlapping at the two largest resolutions ($\bar{S}=10^4$). 
  As before, convergence refers to the $1^{\rm st}$ linear phase only.}
  \label{fig:schemecomparison}
\end{figure}

\begin{figure*}
  \centering
  \includegraphics[width=0.365\textwidth]{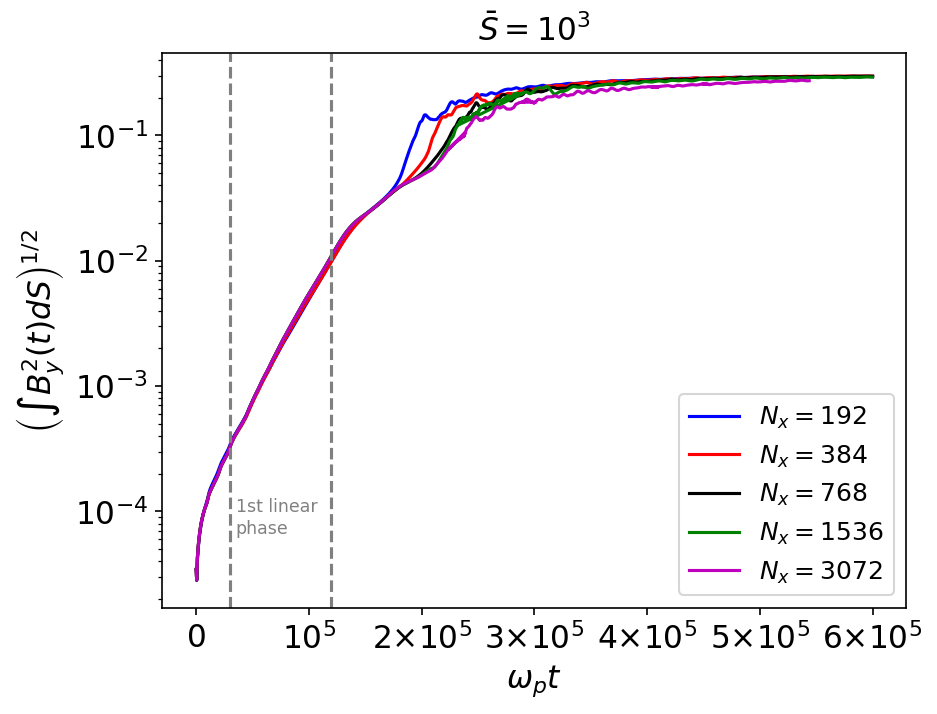} 
  \includegraphics[width=0.312\textwidth]{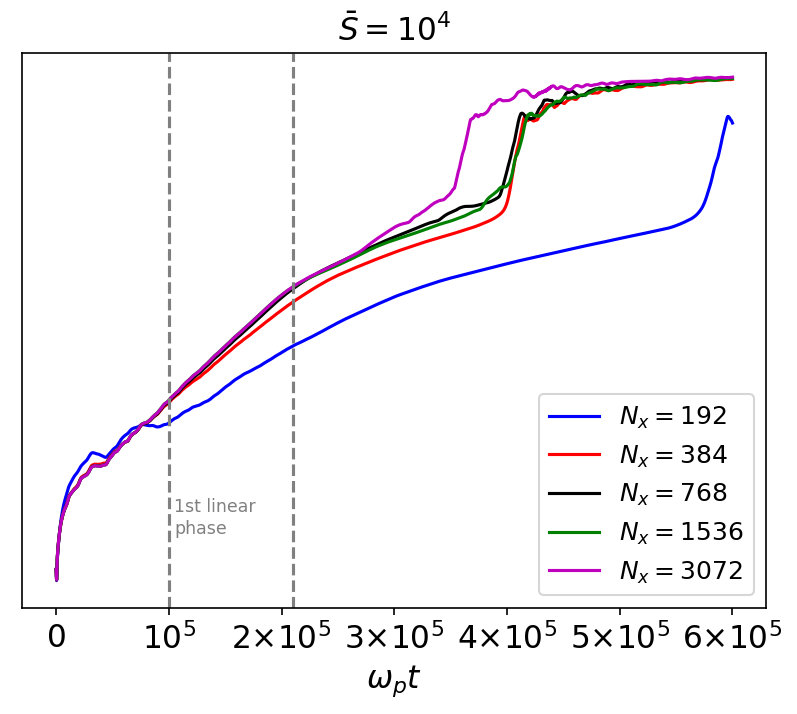}    
  \includegraphics[width=0.307\textwidth]{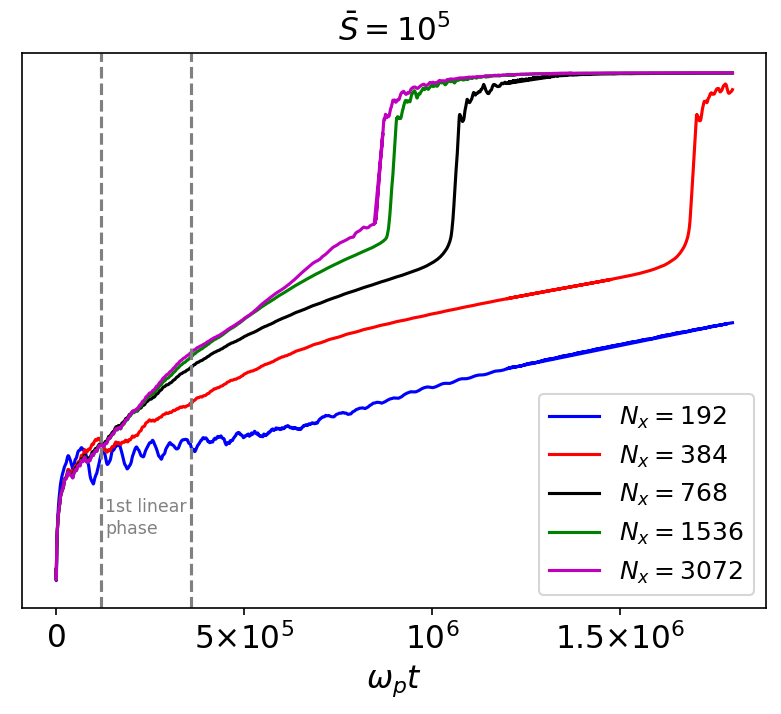}
  \caption{Spatially-averaged transverse component of magnetic field over time for different values of $\bar{S}$ at different resolutions.
  The HLLD scheme with WENO-Z reconstruction has been used. 
  The vertical dashed lines mark the temporal range of the $1^{\rm st}$ linear phase during which convergence is reached.}
  \label{fig:fluid_S}
\end{figure*}

Next, we determine the influence of the Riemann solver as well as the emf-averaging scheme on the computations.
Figure \ref{fig:resvsideal} shows the evolution over time of the spatially-averaged transverse component of magnetic field at different grid resolutions for selected Riemann solvers and emf-averaging schemes (see \S\ref{sec:num_methods}). 
Resistive ($\bar{S} = 10^4$) and ideal ($\bar{S} = \infty$) cases are shown in the upper and lower panels, respectively.

In the resistive case, using the HLL and Roe schemes (top left and middle panels, respectively), the growth rate flattens as the mesh becomes finer and convergence during the $1^{\rm st}$ and $2^{\rm nd}$ linear phases is achieved for $N_x\gtrsim 768$.
Conversely, computations using the HLLD method (top right) reveal a more homogeneous profile where the different evolutionary phases are clearly distinguished at (nearly) all resolutions giving comparable growth and convergence properties. 
This behavior has to be attributed to the reduced amount of numerical dissipation of the HLLD+UCT-HLLD scheme, as discussed in Section 6.2 of \cite{Mignone2021}, for which the diffusive flux terms eventually contributing to the EMF evaluation are proportional to the jump in magnetic fields alone.  

We remind the reader that, in the  ideal case ($\bar{S}=\infty$, bottom panels), the equilibrium condition given by Eq. (\ref{eq:equilibrium}) is a stationary solution of the ideal MHD equations and any dissipative process should be absent. 
In practice, however, the discretization process introduces a numerical viscosity/resistivity  which  allows  the  current sheet to reconnect to some extent. 
Generally speaking, the rate at which field dissipation occurs should depend on the amount of numerical diffusion: more dissipative schemes or lower resolutions will trigger reconnection events earlier.
This is clearly the case for the Roe (+ CT-Contact) and HLL schemes for which convergence will never be reached owing to the resolution-dependent numerical resistivity.
The employment of HLLD scheme reveals, once more, an unexpected benefit: the system remains stable with minimal dissipation at all resolutions and no perturbation growth, as one would expect for an ideal system. 
The same behavior has been witnessed in simulations of ideal current sheets, as described by \cite{Mignone2021} (see Sec. 6.2 of that paper).

This proves that the introduction of a physical resistivity is absolutely essential to ensure convergence with respect to the numerical method and mesh size in simulations of reconnecting current sheets.
This is remarkably true during the linear phase(s) although a word of caution is noteworthy.
For the sake of comparison, in fact, Fig \ref{fig:schemecomparison} plots the spatially-averaged transverse component of magnetic field over time at the two largest resolutions for the selected schemes.
While convergence is reached during the $1^{\rm st}$ linear phase ($\omega_pt\lesssim 2.6\cdot 10^5$), the same does not hold during the subsequent phase where a variation of the magnetic energy is visible depending on the chosen scheme and grid size. 
In fact, during the second phase the width of the current sheet continues to decrease, eventually leading to its fragmentation through the formation of X-points. 
In order to accurately capture this phase, an even larger resolution is needed for resolving the increasingly thinner sheets. 

From these results, we conclude that the HLLD combination scheme with WENO-Z reconstruction seems to produce the most accurate results and it will be employed as our fiducial numerical scheme.

\subsection{Dependence on the Lundquist Number}
\label{sec:Lundquist}
\begin{figure}
  \centering
  \includegraphics[width=0.5\textwidth]{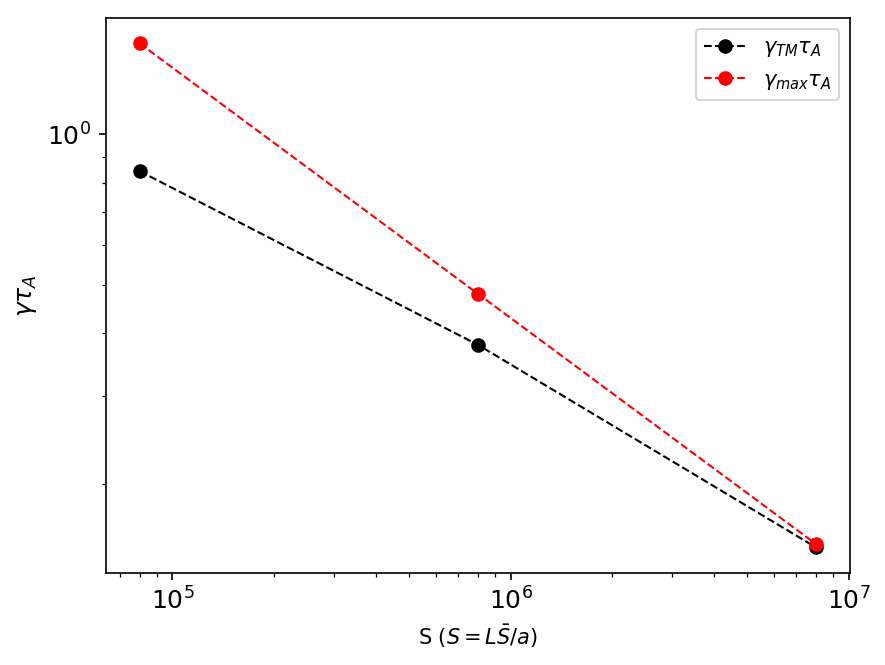} 
  \caption{Comparison between the theoretical growth rate obtained from Eq. (\ref{eq:DelZanna}), in red, and that obtained from the simulations, in black, for different values of $S$.}
  \label{fig:delzanna}
\end{figure}

\begin{figure*}
  \centering
  \includegraphics[width=0.95\textwidth]{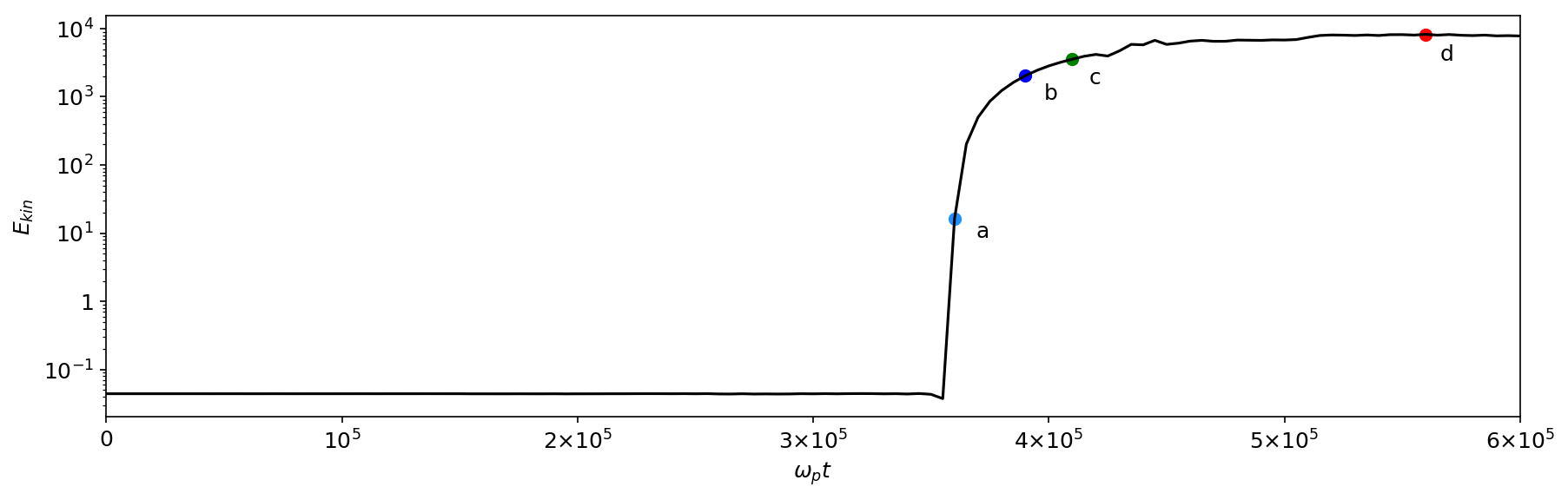} \\
  \includegraphics[width=0.45\textwidth]{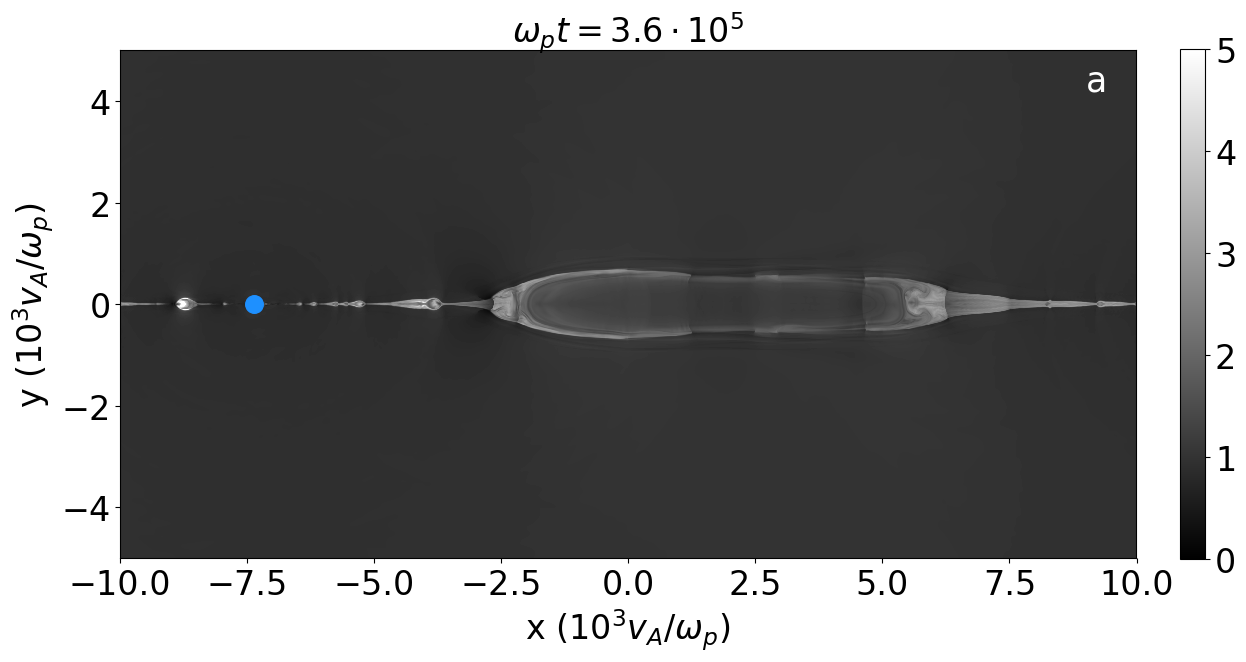} \hspace{0.5pt}
  \includegraphics[width=0.45\textwidth]{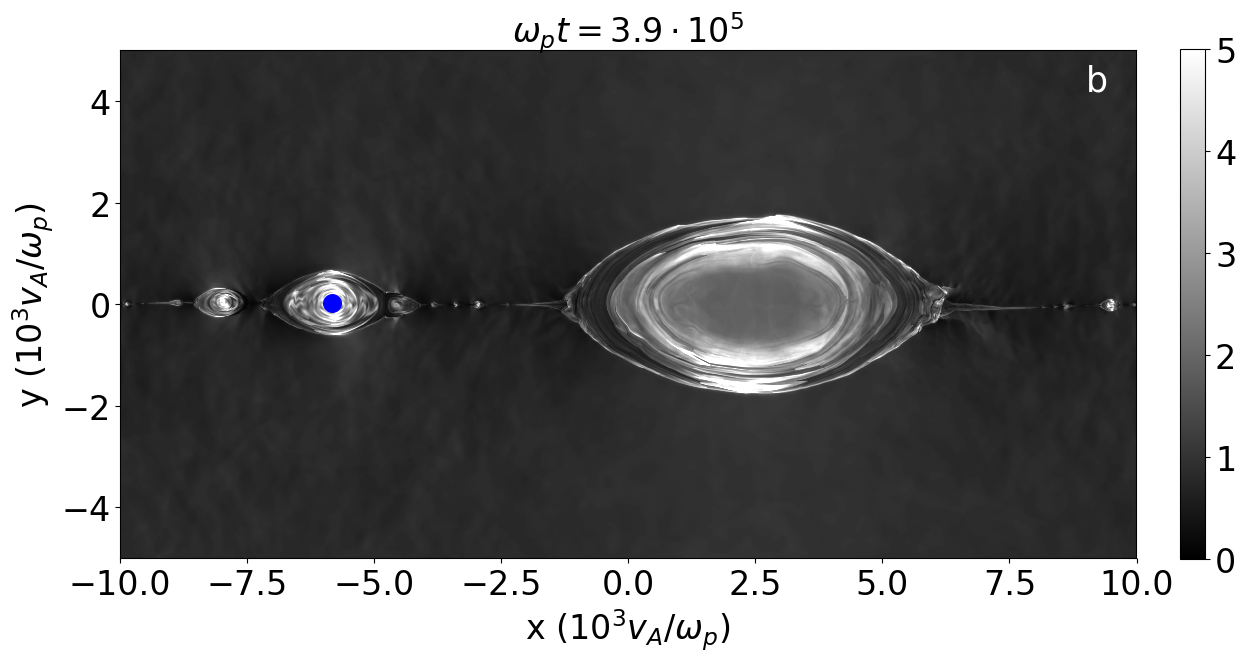} \\
  \includegraphics[width=0.45\textwidth]{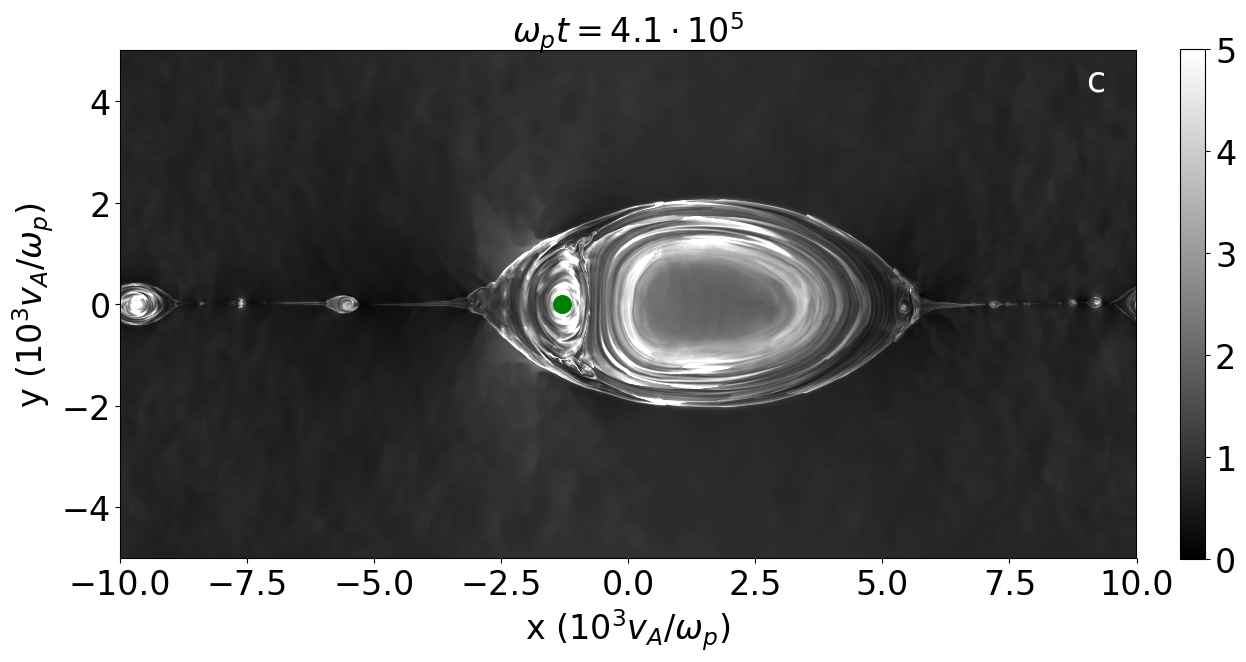} \hspace{0.5pt}
  \includegraphics[width=0.45\textwidth]{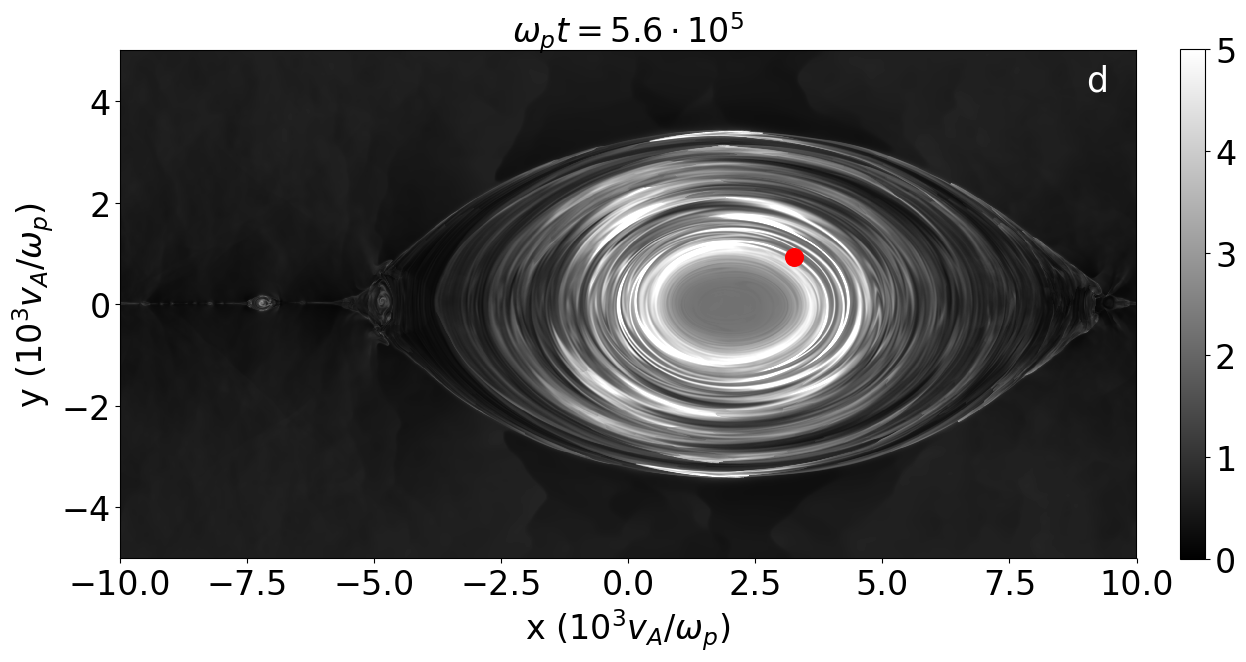} 
  \caption{\textit{Top panel}: Evolution over time of the kinetic energy of the most energetic particle, where the points represent the four instants of time corresponding to the plots in the lower panels.  \textit{Middle and bottom panels}: Overplot of the position of the most energetic particle, colored according to its Lorentz $\gamma$-factor, on the plasma density (colorbar) snapshots in four main moments of its evolution, obtained with a grid resolution $N_x=3072$ and $\bar{S}=10^4$.}
  \label{fig:position2258346}
\end{figure*}

Fig. \ref{fig:fluid_S} shows the spatially-averaged transverse component of magnetic field over time for different values of $\bar{S}$ and grid resolutions.
Note that the saturation phase occurs at later times when $\bar{S}$ increases, i.e. as the physical resistivity decreases.
For this reason, when $\bar{S}=10^5$, the final simulation time has been extended to $\omega_p t_{\rm stop} \approx 1.8 \cdot 10^6$, i.e. three times than for the previous cases.
Our results indicate that the system evolution converges at increasingly larger grid resolution depending, as expected, on the relative magnitude between numerical and physical resistivity, scaling as $\eta_{\rm num}\sim O(\Delta x^2)$ and $\eta \sim 1/\bar{S}$, respectively\footnote{Note that, albeit the employment of fifth-order accurate reconstruction, our computations remain $2^{\rm nd}$-order accurate.}.
As an order of magnitude estimate we expect, therefore, the resolution threshold for convergence to scale approximately as $a/\Delta x \sim 10\sqrt{\bar{S}/10^4}$ for a second-order accurate scheme.
Indeed, for $\bar{S}\approx 10^3$ convergence is observed at all resolutions ($a/\Delta x \gtrsim 3$). 
On the contrary, at $\bar{S}=10^5$, convergence is fully attained only at $N_x\sim3072$ while the low-resolution simulation ($N_x = 192$) shows that the linear growth proceed slowly and the saturation phase is not even reached by the end of the simulation ($\omega_p t_f \approx 1.8 \cdot 10^6$).

It is interesting to compare the measured growth rate for the $1^{\rm st}$  linear phase with the theoretical expectation \citep{DelZanna2016} according to which, in the limit of large $S$, one has
\begin{equation}    \label{eq:DelZanna}
    \gamma_{\max} \tau_A \simeq 0.6 S^{-1/2} (a/L)^{-3/2},
\end{equation}
where $\gamma_{\max}$ is the growth rate of the most unstable mode, $\tau_A = L/v_A$ is the Alfv{\'e}nic time defined on $L$ and $S$ is the Lundqusit number defined according to Eq. (\ref{eq:S}).
Figure \ref{fig:delzanna} shows the growth rate obtained from Eq. (\ref{eq:DelZanna}) compared with that obtained from the simulations (Eq. \ref{eq:slope}) for different values of $S = L\bar{S}/a$ with $\bar{S}=10^3, 10^4, 10^5$.
We find that $\gamma_{\textsc{\tiny{\rm TM}}}\tau_A$ approaches the asymptotic value $\gamma_{\max}\tau_A$ as $S$ increases. 
In fact, for $\bar{S} = 10^5$ we obtain $\gamma_{\textsc{\tiny{\rm TM}}}\tau_A \approx 0.149$ and   $\gamma_{\max}\tau_A \approx 0.152$.


\section{Test Particle Acceleration}
\label{sec:results_part}
%
%

We now turn our attention to the impact of the numerical method and resistivity on the dynamics and energetics of relativistic test-particles which evolve concurrently with the fluid. 

\subsection{Particle Acceleration \& Energetics}
\label{sec:acceleration}
\begin{figure}
  \centering
  \includegraphics[width=0.5\textwidth]{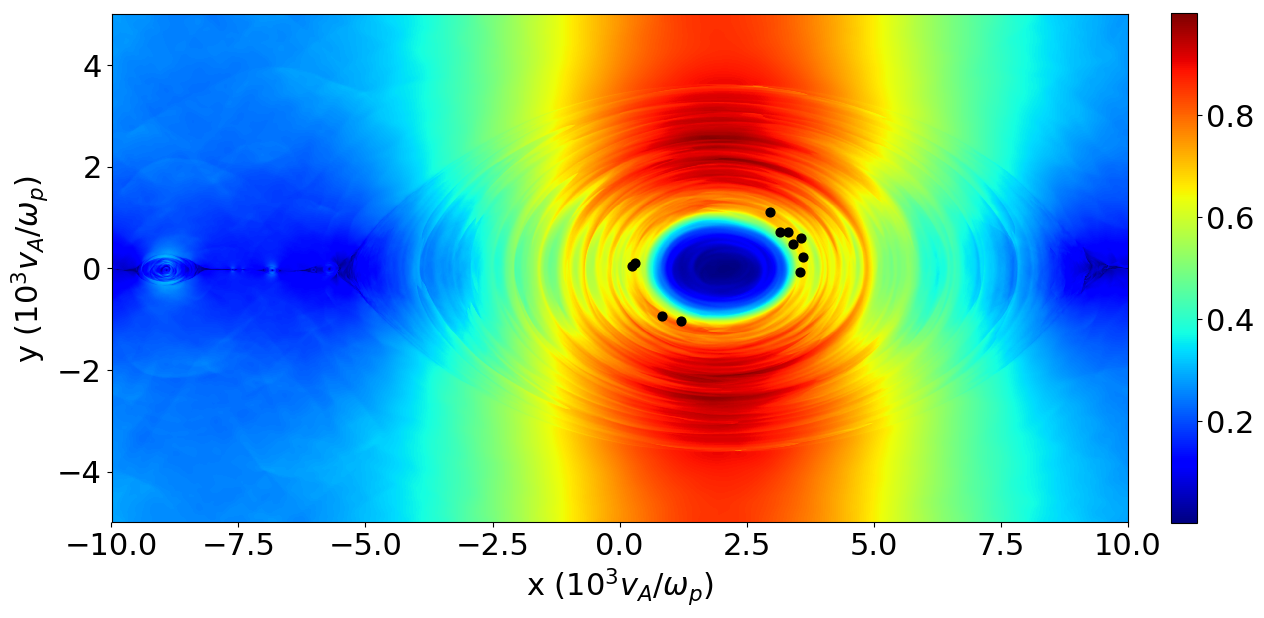}
  \caption{Position of the most energetic particles on the snapshot of the magnetic field module, normalized to its maximum value, at the end of the simulation obtained with a grid resolution $N_x=3072$ and $\bar{S}=10^4$.}
  \label{fig:Bmag3072}
\end{figure}

%
Figure \ref{fig:position2258346} (top panel) shows the kinetic energy history  of the most energetic particle, $E_{\rm kin} = u_p^2/(\gamma + 1)$ where $u_p$ and $\gamma$ represent the module of the particle's four-velocity and its Lorentz factor respectively.
Notice that, for our choice of ${\mathbb C} = 10^4$, particles remain non-relativistic for the entire evolution ($\gamma_{\max} \sim 1$).
A sharp increase is first observed for $ \omega_p t \gtrsim 3.6 \cdot 10^5$, when the particle enters one of the small magnetic islands after crossing an X-point (panel a) where a strong electric field boosts the particle velocity. 
Secondary acceleration events occur and continues concurrently with the process of repeated island merging (panel b) until the merger (panel c) with 
a large final plasmoid, inside which most energetic particles remain trapped (panel d).
Island mergers can lead to an extra particle energy boost, due to the anti-reconnection electric field\footnote{The relative importance of the resistive and convective term of the electric field in Eq. (\ref{eq:electricfield}) in the particle acceleration process will be the subject of a companion paper.} at the secondary current sheet that forms perpendicular to the main one, at the interface between both islands \citep{Oka2010, Sironi2014, Nalewajko2015, Cerutti2019}. 
Once the particle enters the larger magnetic island, it undergoes a sharp acceleration inside the magnetized ring around the plasmoid center within which it is trapped. 
The particle increases its kinetic energy through the $1^{\rm st}$-order Fermi process, since the hosting plasmoid compresses while it merges with smaller islands while accreting particles and magnetic flux \citep{Drake2010, Kowal2011, Guo2014, deGouveia2015, Guo2015, Petropoulou2018, Hakobyan2020}. 
When major mergers no longer occur and the plasmoid stabilizes, the particle energy remains approximately constant in time.
Note that other energetic particles are also subject to the same acceleration mechanisms. 
As an illustrative example, Fig. \ref{fig:Bmag3072} shows a map of the magnetic field module (normalized to its maximum value), together with the position of the most energetic particles at the end of the simulation overplotted. 
As also found by \cite{Petropoulou2018}, this figure indeed highlights that particles dominating the high-energy spectral cutoff reside in a strongly magnetized ring around the plasmoid core.

\begin{figure}
  \centering
  \includegraphics[width=0.48\textwidth]{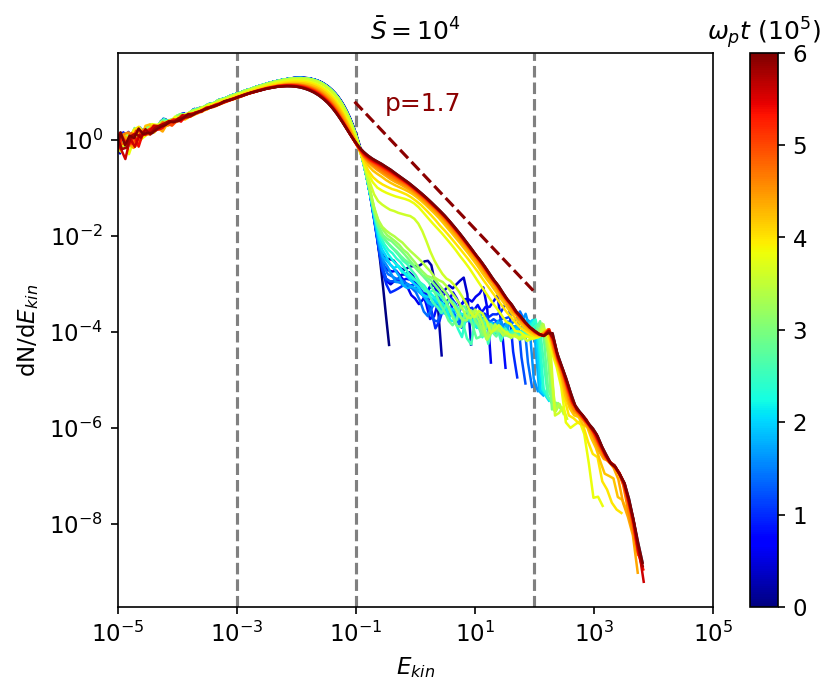}
  \includegraphics[width=0.48\textwidth]{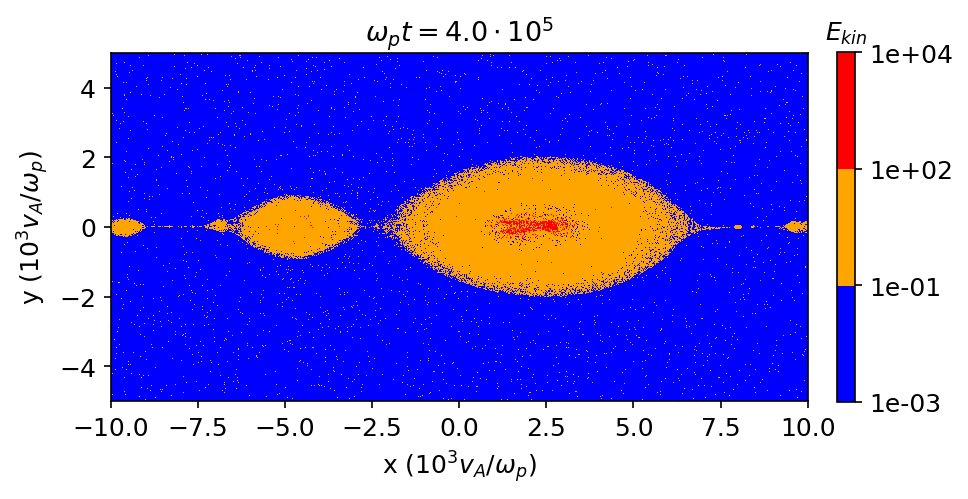}
  \includegraphics[width=0.48\textwidth]{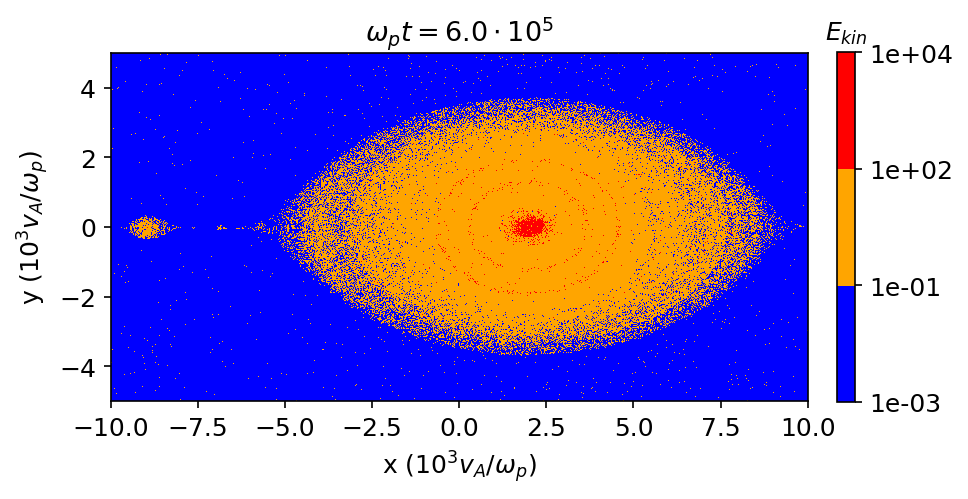}
  \caption{
  Top panel: temporal (colorbar) evolution of particles energy spectrum and the p-index of the power law to which it converges (red dashed line).
  Middle and bottom panel: Spatial distribution of the particles, colored according to their energy (colorbar), at two different instants of time. The energy ranges of the colorbar correspond to the three different parts of the spectra shown in the top panel. 
  The graphs are obtained with a grid resolution $N_x = 3072$ and with $\bar{S} = 10^4$. }
  \label{fig:particles_evolution}
\end{figure}

Figure \ref{fig:particles_evolution} (top panel) shows the temporal evolution of the particles energy distribution (initially set to follow a Maxwellian one).
For convenience, we break the energy range into three portions identified with the low energy end ($10^{-3} \lesssim E_{\rm kin} \lesssim 10^{-1}$), the power-law section $ dN/dE_{\rm kin} \propto E_{\rm kin}^{-p} $ with slope $p \approx 1.7$ ($10^{-1} \lesssim E_{\rm kin} \lesssim 10^2$) and the high-energy cutoff ($10^2 \lesssim E_{\rm kin} \lesssim 10^4$).  
The spectrum reaches an almost asymptotic shape during the saturation phase  when $\omega_p t \gtrsim 4.4 \cdot 10^5$ ($t/ \bar{\tau}_A \gtrsim 1760$, see top right panel of Fig. \ref{fig:resvsideal}).
Note that all spectra shown in this paper are normalized to the total number of particles, which varies with the grid resolution.
Figure \ref{fig:particles_evolution} (middle and bottom panel) shows the particles spatial distribution at $\omega_pt \approx 4\times 10^5$ and $\omega_pt\approx 6\times 10^5$ coloured by the chosen energy ranges.
The low energy end of the spectra (the blue region) is determined by most of the particles in the domain which do not experience significant acceleration.
These particles are predominantly found in the regions outside the plasmoids with spatial and velocity distributions remaining close to the initial values.
Particles populating the power-law component of the spectrum (orange region), on the contrary, are found in proximity of the current sheet or settle in the outermost rings of the magnetic islands.
Their number increase in time as the acceleration mechanism produces more energetic particles.
Particles with the highest energy approximately, (red regions) are settled inside the current sheet as reconnection begins and then gradually fill the central regions of the plasmoids as they form and merge with each other.
At the end of the computation (bottom panel) we find that approximately $\sim 51\%$ of particles fill the low-energy domain, $\sim 47\%$ particles populate the power-law section of the spectrum while $\sim 2\%$ is represented by high-energy particles.

\subsection{Dependence on Grid Resolution and Numerical Method}
%

\begin{figure*}
  \centering
  \includegraphics[width=0.515\textwidth]{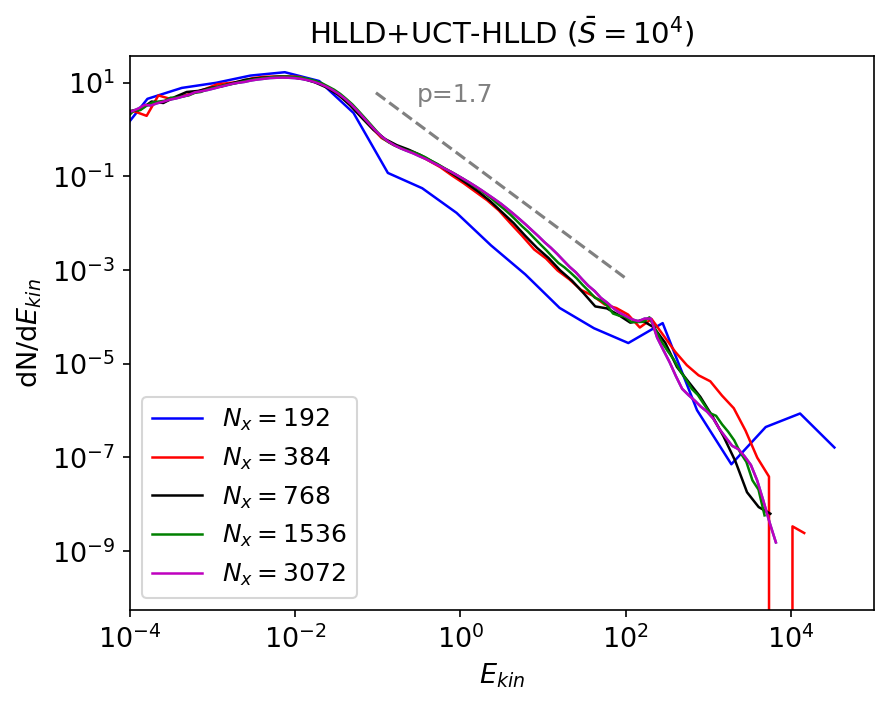} \hspace{0.2pt}
  \includegraphics[width=0.47\textwidth]{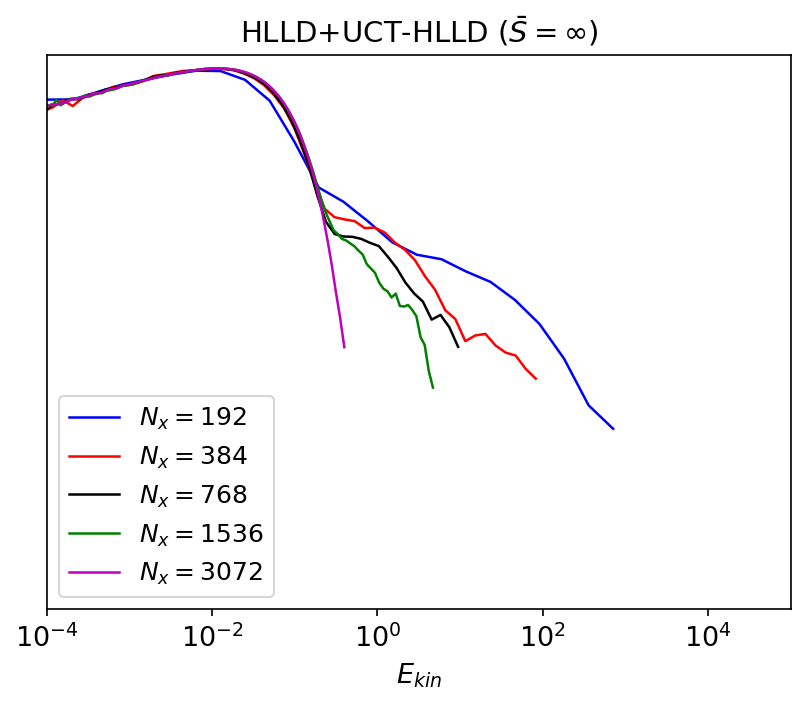} \\
    \includegraphics[width=0.515\textwidth]{{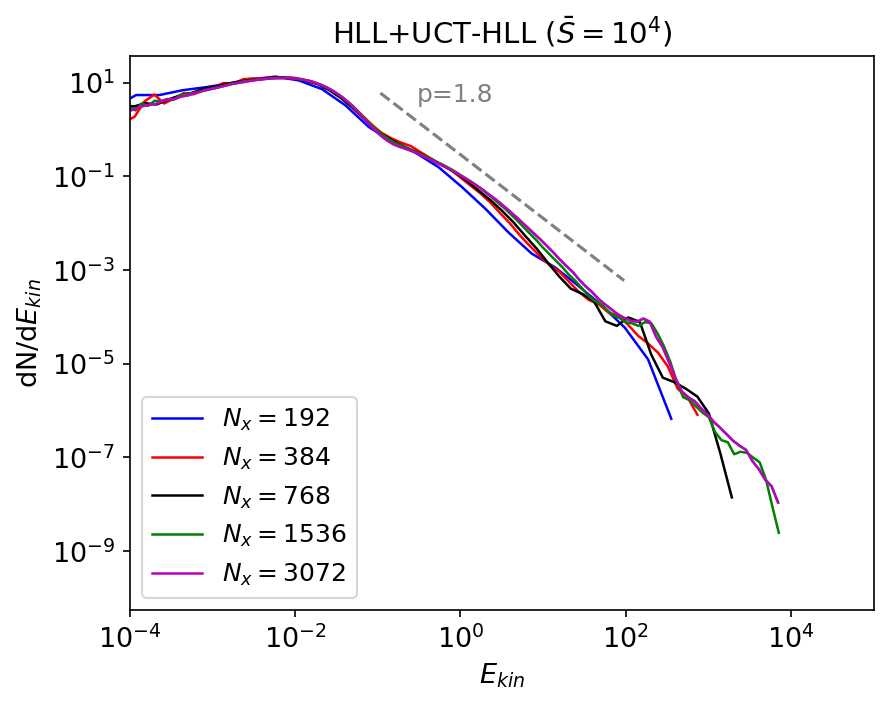}} \hspace{0.2pt}
  \includegraphics[width=0.47\textwidth]{{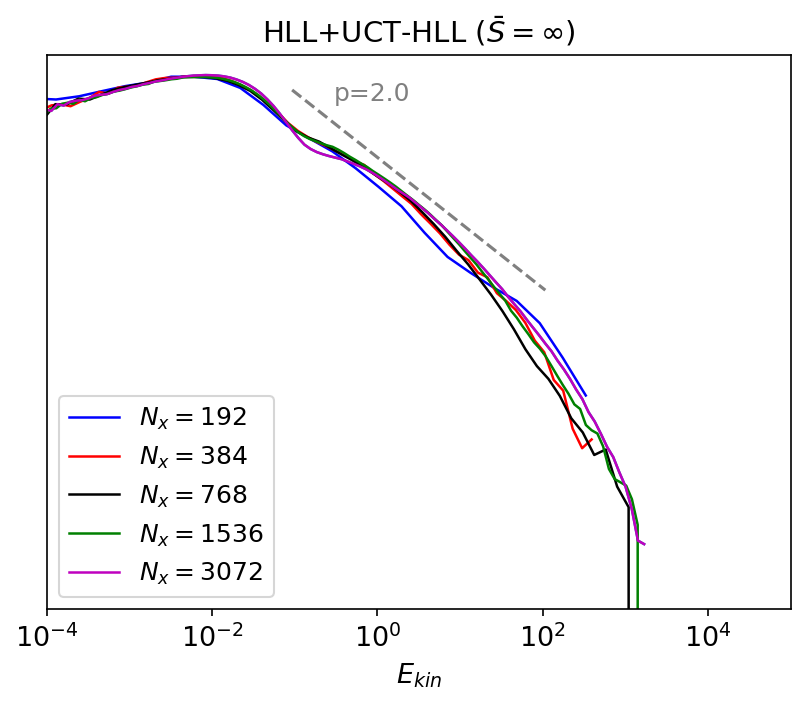}}
  \caption{Comparison of particle energy spectra at the end of the simulation ($\omega_p t=6 \cdot 10^5$) for different grid resolutions in the case of a resistive with $\bar{S}=10^4$ (left panels) and an ideal (right panels) plasma. The HLLD (+ UCT-HLLD) scheme is used in the upper panels, with the HLL (+ UCT-HLL) one is used in the lower panels. 
  The dashed gray lines represents the power law to which the spectra converge, with the corresponding p-index. 
  In all the cases the WENO-Z reconstruction has been used.
    }
  \label{fig:spectraconvergence}
\end{figure*}

\begin{table}
	\centering
	\caption{$p$-index of the power-law part of the spectra (referring to Fig. \ref{fig:spectraconvergence}) at different resolutions for different combinations of schemes in the case of resistive ($\bar{S}=10^4$) and ideal ($\bar{S}=\infty$) plasma. The ideal HLLD+UCT-HLLD case is not shown as magnetic reconnection does not start.}
	\label{tab:pindex}
    \centering
	\begin{tabular}{ccccc} 
                \hline
                & & \multicolumn{3}{c}{Power-law index p}  \\
                \cmidrule{3-5}
                Resolution & $a/\Delta x$ & HLLD & HLL & HLL  \\
                &  & ($\bar{S}=10^4$)   &  ($\bar{S}=10^4$)   & ($\bar{S}=\infty$) \\
		\hline
		192 $\times$ 96 & $\sim2.5$ & 1.5 & 1.3 & 1.5  \\
		384 $\times$ 192 & $\sim5$ & 1.6 & 1.7 & 2.2 \\
		768 $\times$ 384 & $\sim10$ & 1.7 & 1.8 & 2.4  \\
                1536 $\times$ 768 & $\sim20$ & 1.7 & 1.8 & 2.2 \\
                3072 $\times$ 1536 & $\sim40$ & 1.7 & 1.8 & 2.0 \\
		\hline
	\end{tabular}
\end{table}

We now assess the impact of grid resolution and physical resistivity on the particle energy distribution.

A comparison between particles energy spectra at the end of the simulation ($\omega_p t \approx 6 \cdot 10^5$) at different resolutions is shown in the left and right panels of Figure \ref{fig:spectraconvergence}, in the resistive ($\bar{S}=10^4$) and ideal ($\bar{S}=\infty$) simulation cases, for the most and least diffusive numerical schemes, HLL (lower panels) and HLLD (upper panels), respectively. 
Table \ref{tab:pindex} shows the corresponding spectral index $p$ of the power-law part of the spectrum.

In the presence of a physical resistivity (left panels), the spectrum remains almost unchanged once $N_x \gtrsim 768$ ($a/\Delta_x \simeq 10$) and it quickly converges to a power-law with index $1.8\lesssim p \lesssim 1.7$ for both numerical methods, as shown in the first column of Table \ref{tab:pindex}, indicating that once the tearing instability is triggered and the reconnection cascade commences, the acceleration properties are virtually independent of the numerical resolution and numerical diffusion.

However, if we consider an ideal plasma (right panels), the results differ depending on the chosen numerical method.
For the HLLD scheme, which has least dissipation, the tearing instability is gradually quenched (as already discussed in \S\ref{sec:riemann}) as the resolution increases and no significant particle acceleration takes place so that the particle energy distribution remains close to the initial Maxwellian (top right in Fig. \ref{fig:spectraconvergence}).
Conversely, the presence of a larger numerical diffusion in the HLL scheme, triggers magnetic reconnection even in the ideal limit, thus spawning a spectral distribution with power law index $p \approx 2.0$ (see the last column of Table \ref{tab:pindex}).
We remind the reader that, in this case, fluid convergence (in the sense discussed in \S\ref{sec:riemann}) could not be achieved (see Fig. \ref{fig:resvsideal}).

\begin{figure*}
  \centering
  \includegraphics[width=0.35\textwidth]{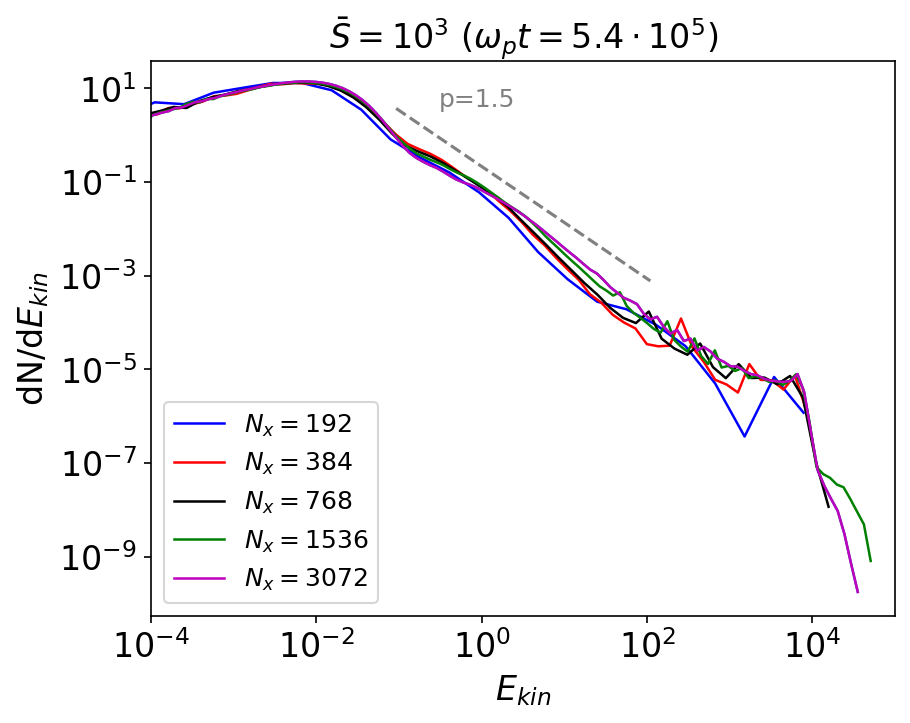} \hspace{0.2pt}
  \includegraphics[width=0.315\textwidth]{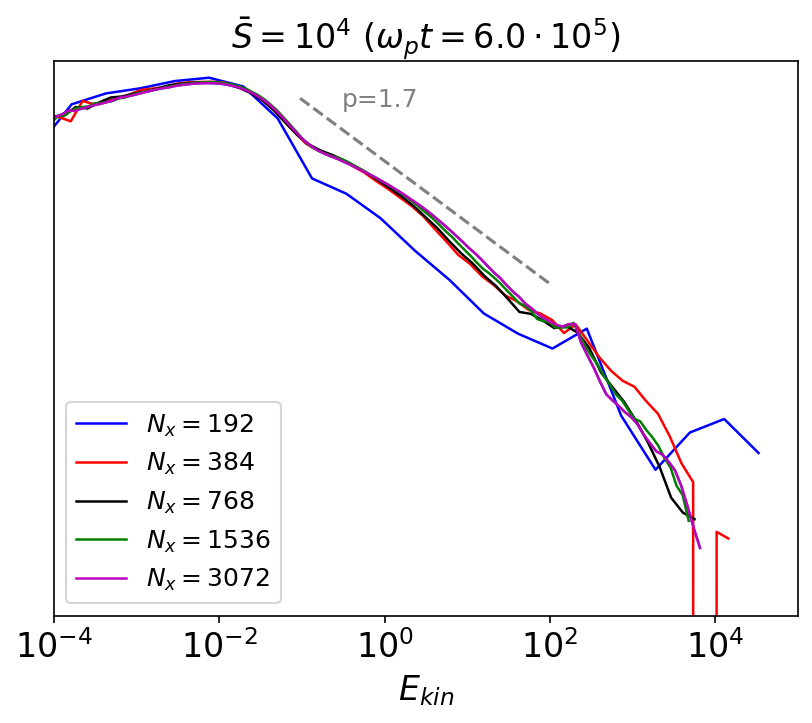} \hspace{0.2pt}
  \includegraphics[width=0.315\textwidth]{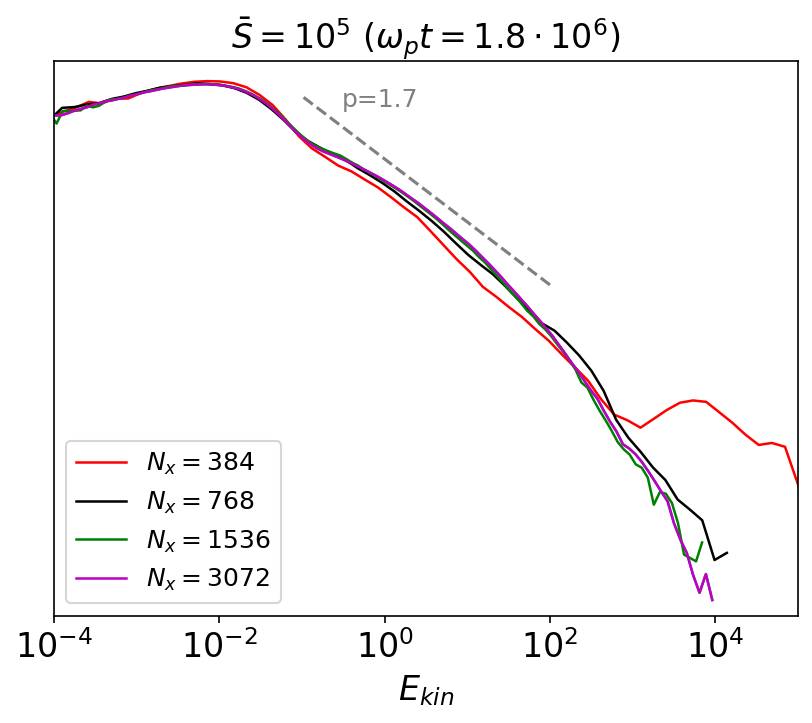}
  \caption{Comparison of particle energy spectra in the saturation phase at different $\bar{S}$ for several grid resolutions, obtained using the HLLD+UCT-HLLD with the WENO-Z scheme.}
  \label{fig:spectraScomparison}
\end{figure*}

While the spectral index obtained in the ideal case is not significantly different from the resistive case, our results indicate that the outcome of ideal MHD computations should be interpreted with some caution as the mechanisms triggering resistive instability may be driven in a rather unpredictable way by numerical diffusion rather than by actual physical effects. This conclusion may differ if we consider the case of fast reconnection driven by turbulence.

\subsection{Spectral Distribution vs Lundquist Number}
%

Next, we repeated the resolution study by varying the value of the redefined Lundquist number $\bar{S}$. 
Figure \ref{fig:spectraScomparison} shows a comparison between the particle energy spectra during the saturation phase, for different values of $\bar{S}$ and grid resolutions. 
As shown in Figure \ref{fig:fluid_S}, the beginning of the saturation phase is delayed as $\bar{S}$ increases, for this reason the spectra shown in Fig. \ref{fig:spectraScomparison} correspond to different computational times.
Also, for higher value of the Lundquist number ($\bar{S}=10^5$) we omit results from low resolution simulation ($N_x = 192$, i.e. $a/\Delta x \simeq 2.5$), since, as mentioned in the section \ref{sec:riemann}, magnetic reconnection at that time has just started.

From the figure it appears that the power-law index weakly depends on the Lundquist number and it converges to a value $1.5\lesssim p \lesssim 1.7$ again for grid resolutions at around $N_x \gtrsim 768$ ($a/\Delta x \simeq 10$).
A quantitative measure of the $p$-index is provided, for different values of $\bar{S}$, in Table \ref{tab:pindex2} at the largest grid resolution ($N_x = 3072$, i.e. $a/\Delta x \simeq 40$). 
Note that it remains substantially the same for $\bar{S} \gtrsim 10^4$ (or $S \gtrsim 8 \cdot 10^5$), a result that may be connected to the onset of the fast reconnection regime \citep{Landi2015}. 


These results lead us to conclude that, once magnetic reconnection has started, the amount of resistivity - either of numerical or physical origin - has a weak (or almost negligible) impact on the particle energization process.
Similar findings have been established, for instance, for particle acceleration in MHD turbulence-induced fast reconnection systems with different driving processes of turbulence \citep[see, e.g.,][]{Kowal2012, delValle2016, Medina2021}.

\begin{table}
	\centering
	\caption{p-index of the power-law part of the spectra (referring to Fig. \ref{fig:spectraScomparison}) at at the highest grid resolution ($N_x = 3072$) for different values of $\bar{S}$.}
	\label{tab:pindex2}
    \centering
	\begin{tabular}{cccccc} 
                \hline
                & & \multicolumn{4}{c}{Power-law index p}  \\
                \cmidrule{3-6}
                Resolution & $a/\Delta x$ & $\bar{S}=10^3$ & $\bar{S}=10^4$ & $\bar{S}=10^5$ &
                $\bar{S}=\infty$ \\
		\hline
                3072 $\times$ 1536 & $\sim40$ & 1.5 & 1.7 & 1.7 & - \\
		\hline
	\end{tabular}
\end{table}

Our results favourably compare to previous works using test-particles in MHD snapshots, in particular with \cite{Gordovskyy2010a,Gordovskyy2010b} who found a power law index $p \approx 1.5-3.0$ at the end of reconnection (O-point stage).
In addition, PIC simulations of merging plasmoids reveal a spectral index compatible with our results, that is $p \approx 1.5$ \citep[see, for instance,][]{Drake2010, Drake2013,deGouveia2015}.

\section{Summary and discussion}
\label{sec:summary}
 
In this work we have presented high-resolution 2D numerical simulations of tearing-unstable MHD current sheet embedding a non-thermal population of test-particles evolving along with the fluid.
The initial condition for the thermal plasma consists of a pressure-balanced magnetized ($\beta = 0.01$) Harris current sheet with constant resistivity.
Computations are performed using the PLUTO code for different Lundquist numbers. 
We do not limit particle integration to frozen MHD snapshots  \citep[as done, for instance, by][]{Liu2009, Gordovskyy2010a, Gordovskyy2010b, Kowal2011, Ripperda2017a} but, rather, they account for the concurrent evolution of both fluid \emph{and} particles (without mutual feedback) similarly to \cite{Gordovskyy2010b} or \cite{Ripperda2017b}.

Our goal aimed at quantifying the impact of the numerical method, grid resolution and physical resistivity on both i) the current-sheet evolution and its convergence properties as well as ii) on the spectral properties of non-thermal test-particles evolving within the background thermal plasma. 

In the first part of the paper, we initially focused only on the fluid evolution, identifying four main temporal phases characterized by a growing number of newly forming X-points.
After i) an initial $1^{\rm st}$ linear phase characterized by a shrinking of the initial current sheet ($t/\bar{\tau}_A \lesssim 840$), ii) a $2^{\rm nd}$ linear phase (ending at $t/\bar{\tau}_A \approx 1360$), marks the evolution of smaller current sheets, resulting from the breaking of the initial one,; iii) a more rapid fragmentation phase leads to the appearance of several X- and O-points feeding the formation of dynamically interacting plasmoids ($t/\bar{\tau}_A \lesssim 1840$) and iv) a final nonlinear saturated phase is accompanied by the presence of one large magnetic island. 

Several simulations using different numerical methods and mesh resolutions have demonstrated that convergence during the initial linear stages of the evolution can be achieved only for finite values of the Lundquist number $\bar{S} = av_A/\eta$, where $a$, $v_A$ and $\eta$ are, respectively, the initial current sheet width, Alfv{\'e}n velocity and physical resistivity.
The minimum resolution at which convergence is attained depends on the amount of numerical diffusion inherited from the underlying discretization method.
Below this resolution, the linear growth phase is dominated by spurious numerical effects which, as a general trend, are likely to delay the onset of the tearing instability as the resolution become coarser.
In this respect, we have found that the combination of the HLLD Riemann solver and the UCT-HLLD emf averaging scheme of \cite{Mignone2021}, together with $5^{\rm th}$-order WENO-Z reconstruction, yields the best performance achieving convergence already at $a/\Delta x \simeq 10$ when $\bar{S} = 10^4$.
This is about half the grid resolution when compared to either linear reconstruction or more diffusive numerical methods based on more approximate Riemann solvers \citep[e.g., HLL, see][and reference therein]{delZanna_etal2007} or second-order emf averaging schemes \citep[e.g., CT-Contact, see][]{Gardiner2005} for which convergence is ensured when $a/\Delta x \gtrsim 20$.

For larger (smaller) values of the Lundquist number the mesh size has to be increased (decreased) at the point where numerical diffusion falls below the physical one.
For a globally second-order accurate scheme we have shown that this is expected to hold if the number of computational zones covering the initial current sheet width scales approximately as $a/\Delta x \sim 10\sqrt{\bar{S}/10^4}$.
We also have verified that the linear growth rate matches the theoretical prediction for asymptotically large $\bar{S}$.
Conversely in the ideal case ($\bar{S}=\infty$), we have observed that the discretization scheme introduces a grid-dependent numerical resistivity that still allows the current sheet to reconnect, although convergence can never be actually achieved. 
This is easily explained by the fact that a change in grid resolution tantamount to a different problem with another value value of the (spurious) resistivity. 
Only with the employment of the HLLD+UCT-HLLD the system remained stable as one would expect for an ideal current sheet.
Based on this results, we have picked the HLLD scheme (with fifth-order WENO-Z spatial reconstruction) as our optimal numerical method.
Future studies will probably take advantage of genuine $4^{\rm th}$-order schemes.

Once the system enters in its nonlinear stage, the spatially-averaged transverse component of the magnetic field reaches the same value regardless of the Lundquist number. This result is analogous to the turbulent situation  \cite{Lazarian1999, Kowal2009}, in which the properties of the system do not depend on the Lundquist number but only on the properties of turbulence as long as the inertial range is captured.

In the second part of this work we have examined how tearing-unstable current sheets become favourable to particle acceleration and energization.
Particles that reach the highest energies at the end of the simulation share the same acceleration mechanisms. 
They initially cross an X-point and enter inside a small plasmoid that begins to merge with the adjacent ones.
During the merger process, the anti-reconnection electric field brought at secondary current sheets from merging plasmoids \citep{Oka2010, Sironi2014, Nalewajko2015, Cerutti2019} is responsible for the first acceleration step.
Particles eventually remain trapped inside plasmoid mergers and continue to gain energy through a $1^{\rm st}$-order Fermi mechanism \citep{Drake2010, Kowal2011, Guo2014, deGouveia2015, Guo2015, Petropoulou2018, Hakobyan2020}. 
As also found by \cite{Petropoulou2018}, the most energetic particles at the end of the simulation ($10^2 \lesssim E_{\rm kin} \lesssim 10^4$) reside in a strongly magnetized ring around the plasmoid core. 
On the contrary, the low-energy particles ($10^{-3} \lesssim E_{\rm kin} \lesssim 10^{-1}$) are predominantly found in the regions outside the plasmoid, and they did not experience significant acceleration. 
In between, particles populating the power-law component of the spectrum ($10^{-1} \lesssim E_{\rm kin} \lesssim 10^2$) are found in proximity of the current sheet or settle in the outermost rings of the plasmoid. 

Several computations at different grid resolutions indicate that the particle energy distribution remains almost unchanged for $a/\Delta x \gtrsim 10$ and it quickly converges to a power law with index $\approx 1.7$, when $\bar{S} \gtrsim 10^4$.
Different values of the Lundquist number, in fact, appear to have a weak influence on the power law index, once the fast reconnection regime ($\bar{S} \gtrsim 10^4$) has been reached.
These results do not generally depend on the integration method or  its numerical diffusion but seem to have a general validity inasmuch the magnetic reconnection process is operating.
Indeed, we have found that this holds even for ideal MHD ($\bar{S}=\infty$, albeit with a different spectral index) for which the island formation process, when present, could be triggered solely by numerical resistivity.
This has been clearly observed in the presence of more dissipative scheme such as the HLL Riemann solver (for which 
$p \approx 2$), but it does not appear with the more accurate HLLD Riemann solver / emf averaging combination.

Our conclusion is that, in the context of reconnection-driven test-particle acceleration, there is no need to reach very high grid resolutions and that the amount of resistivity has very weak or almost negligible impact on the particle energization process.
Note that our results are compatible with several studies of turbulence-driven fast reconnection with or without explicit resistivity, see, e.g., \cite{Kowal2011, Kowal2012, delValle2016, Medina2021}.


In a companion paper we will consider the extension to the relativistic (fluid) case as well as the impact of the guide field  \citep[which is still a matter of debate, see, e.g.,][]{Drake2010, Kowal2011} as well as of a static background magnetic configuration (i.e., particles evolving on fluid snapshots) on the particle acceleration process. 
In addition, the relative importance of the advective and resistive electric fields in the particle energization will be thoroughly discussed.


\section*{Acknowledgements}
We acknowledge support by CINECA through the  Accordo Quadro INAF-CINECA for the availability of high performance computing resources (project account INA20\_C6A50). 
We wish to thank Luca Del Zanna and Lorenzo Sironi for valuable comments that improved the quality on this manuscript.

\section*{Data Availability}
PLUTO is publicly available and the simulation data will be shared on reasonable request to the corresponding author.




\bibliographystyle{mnras}
\bibliography{paper} 



\appendix

\section{X-points number algorithm}
\label{app:xpoints}
We illustrate the algorithm employed to locate X-point from our simulation results.

Since, at an X-point, the magnetic field $|\vec{B}|$ vanishes (and so do the $B_x$ and $B_y$ components), we first identify computational zones hosting a local minimum of $|\vec{B}|$ over a stencil of $3\times3$ zones.
These zones, therefore, may potentially contain a null point.
Let $(i,j)$ be the indices of a zone hosting a local minimum of $|\vec{B}|$. 
A bilinear interpolation is used to represent the $B_x$ and $B_y$ components of magnetic field inside a square delimited by the four corner points $(x_{i\pm1},y_{j\pm1})$:
\begin{equation}\label{eq:H(x,y)}
\begin{split}
    H(\hat{x},\hat{y}) &=    H_{i-1,j-1}(1 - \hat{x})(1 - \hat{y}) 
               + H_{i+1,j-1} \hat{x} (1 - \hat{y}) + \\
           &   + H_{i-1,j+1}(1 - \hat{x})\hat{y} 
               + H_{i+1,j+1}\hat{x}\hat{y} ,
\end{split}
\end{equation}
where $H(\hat{x},\hat{y})$ denotes either the $x$- or $y$-component of $\vec{B}$ while $\hat{x}$ and $\hat{y}$ are normalized coordinate in $[0,1]$.  

We then require that both $B_x(x,y)$ and $B_y(x,y)$ have a root,
\begin{equation}\label{eq:bilinear_sys}
\begin{cases}
  B_x(\hat{x},\hat{y}) = a_0 + a_1 \hat{x} + a_2 \hat{y} + a_3 \hat{x}\hat{y} = 0 \\
  B_y(\hat{x},\hat{y}) = b_0 + b_1 \hat{x} + b_2 \hat{y} + b_3 \hat{x}\hat{y} = 0,
\end{cases}
\end{equation}
where the coefficients $a_0, a_1, ..., b_3$ are readily found from Eq. (\ref{eq:H(x,y)}).
Eq. (\ref{eq:bilinear_sys}) leads to a quadratic equation whose solutions are considered null points only if they fall inside the unit square.


\bsp	
\label{lastpage}
\end{document}